\documentstyle[12pt]{article}
\begin{document}
\baselineskip 0.71cm
\newcommand{\tri}{\triangleright}
\newcommand{\range}{{\rm range}}
\newcommand{\Ree}{{\rm Re }}
\newcommand{\Imm}{{\rm Im }}
\newcommand{\diag}{{\rm diag}}
\newcommand{\sign}{{\rm sign}}
\newcommand{\tr}{{\rm tr}}
\newcommand{\rank}{{\rm rank}}
\newcommand{\bp}{\bigskip}
\newcommand{\mdp}{\medskip}
\newcommand{\slp}{\smallskip}
\newcommand{\Rw}{\Rightarrow}
\newcommand{\ts}{& \hspace{-0.1in}}
\newcommand{\nn}{\nonumber}
\newcommand{\bea}{\begin{eqnarray}}
\newcommand{\eea}{\end{eqnarray}}
\newcommand{\beas}{\begin{eqnarray*}}
\newcommand{\eeas}{\end{eqnarray*}}
\newcommand{\beq}{\begin{equation}}
\newcommand{\eeq}{\end{equation}}
\newtheorem{exa}{Example}[section]
\newtheorem{thm}{Theorem}[section]
\newtheorem{lem}{Lemma}[section]
\newtheorem{prop}{Proposition}[section]
\newtheorem{fact}{Fact}[section]
\newtheorem{cor}{Corollary}[section]
\newtheorem{defn}{Definition}[section]
\newtheorem{rem}{Remark}[section]
\renewcommand{\theequation}{\thesection.\arabic{equation}}
\title{Koopman System Approximation Based Optimal Control of Multiple
  Robots---Part I: Concepts and Formulations}
\author{{\it Gang Tao} and {\it Qianhong Zhao} \\
Department of Electrical and Computer Engineering\\
University of Virginia\\
Charlottesville, VA 22904}
\date{}
\maketitle

\begin{abstract}
This paper presents a study of the Koopman operator theory and its
application to optimal control of a multi-robot system. The Koopman
operator, while operating on a set of observation functions of the
state vector of a nonlinear system, produces a set of dynamic
equations which, through a dynamic transformation, form a new dynamic
system. As an operator, it has a rich spectrum of mathematical
properties, and as a tool for dynamic system analysis and control
design, it has a unique collection of practical meanings.

\medskip
The Koopman system technique is then applied to the 
development of a linear or bilinear model approximation of nonlinear
utility functions for optimal control of a system of multiple (mobile)
robots, by selecting the utility functions as the Koopman system state
variables and expressing the set of Koopman variables as the state
variables of a linear or bilinear system whose parameters are
determined through optimization. An iterative (online) algorithm is 
developed for adaptively estimating the parameters of the
approximation model of the robot system with nonlinear utility
functions.

\medskip
Finally, the control problems based on a linear or bilinear model for
the nonlinear utility functions are formulated for optimal control of
the multi-robot system, by transforming the nonlinear programming
problem to a linear programming problem.
\end{abstract}

\begin{quote}

\end{quote}

\setcounter{equation}{0}
\section{Introduction}
Recently, research on the Koopman systems induced by the Koopman
operator \cite{k31} has shown broad and significant progresses in
both theory and applications, and there is a vast amount of literature
on the related topics, see, for example, \cite{lm13}-\cite{bfv21}
addressing some topics such as definitions, theory and system
applications of Koopman operators. This paper uses the basic concepts
and theory developed in these references, and develops some extensions
and application to the multi-robot system with nonlinear utility
functions to be approximated by a linear model seen in the literature
and a bilinear model derived in the paper.

The Koopman operator (a transformation defined in \cite{k31}), when
applied on a set of observation functions (observables) $\psi(x(k))$
of the state vector of a (discrete-time) nonlinear dynamic system
$x(k+1) = f(x(k), u(k))$, produces the future values $\psi(x(k+1))$
(of the observables) which are equal to $\psi(f(x(k), u(k)))$ because
of $x(k+1) = f(x(k), u(k))$. Thus, a dynamic relationship is
generated, which, via a functional transformation, induces a new
dynamic system with $\psi(x(k))$ as its state variables, which, when
properly constructed and approximated, can facilitate the analysis and
feedback control design for the original nonlinear system. 

In theory, under the Koopman operation and a necessary functional
transformation, a set of chosen system state observation functions
lead to a well-defined Koopman system, and certain choices of
observation functions can lead to some canonical form Koopman
system. In practice, some performance-related functions may need to be
selected as the observation functions (the Koopman variables) and the
associated subsystem dynamic functions can be nonlinear and
noncanonical. Those functions should be approximated well for helping
to solve the related control problems, as a main motivation of using a
Koopman system formulation. For many applications, it is often crucial
to go beyond a linear model approximation of the resulted Koopman
system, to use, for example, a bilinear model, in order to increase
the model accuracy for improved system performance.

This paper pursues some related issues of Koopman
systems. Theoretically, the relationship $\psi(x(k+1)) = \psi(f(x(k),
u(k)))$ from the Koopman operation is not yet a formal dynamic
equation (a Koopman system equation, as the base for applications of
Koopman systems, such as linear or bilinear approximations of a
nonlinear Koopman system equation) which should contain $\psi(x(k))$
(in stead of $x(k)$) in $\psi(f(x(k), u(k)))$. A formal functional
transformation is needed to construct such a dynamic
equation. Practically, optimal control or model predictive control
with nonlinear utility (performance) functions (to be maximized by
feedback control) usually needs to solve some nonlinear programming
problems which can be computationally overloading. Transformation of
such a nonlinear programming problem to a linear one needs to be
demonstrated. In this paper, such issues will be further considered and
studied.

\medskip
The goals of this paper are to show some of our new research
progresses in this direction:

\begin{quote}
$\bullet$ to establish a functional transformation  to link
  the (discrete-time) Koopman equation ${\cal K}[\psi(x(k))] =
  \psi(x(k+)) = \psi(f(x(k), u(k)))$
  to a dynamic system equation $\psi(x(k+)) = \xi(\psi(x(k)), u(k))$, i.e.,
  $z(k+1) = \xi(z(k), u(k))$ which provides the base for a linear
  approximation model $z(k+1) = A z(k) + B u(k)$ or other dynamic models;

$\bullet$ to develop, based on the established functional 
  transformation, a linear approximation Koopman system model and a
  bilinear approximation Koopman system model for a multi-robot system
  with nonlinear utility functions (the performance indexes to be
  maximized by feedback control), and design iterative (adaptive)
  algorithms for updating the model parameters online with new
  measurements (incoming data); and

$\bullet$ to transform the nonlinear programming problem from optimal
  control of a multi-robot system with nonlinear utility (performance)
  functions, to a linear programming problem, based on the developed
  linear or bilinear Koopman system model.
\end{quote}

With the presentation of the linear model of a Koopman system, we
demonstrate the basic procedure of an approximation model derivation 
and evaluation (for both a linear model and a bilinear model approximations). With
the development of the bilinear model, we expand
the approximation model with additional terms which capture more
dynamics of a nonlinear Koopman system model, to improve the accuracy
of an approximation model.

\medskip
In Section 2, we present some basic concepts of the Koopman operator
and its induced dynamic system (a Koopman system), and derive a
proposition to show the existence of such a system. We also show some
basic ideas of approximating a nonlinear Koopman system by a linear or
bilinear system model. In Section 3, we develop a Koopman system
framework for a multi-robot system with nonlinear utility
(performance) functions for optimal control, which consists of the
construction of a Koopman system with nonlinear utility functions as a
set of observation functions, the development of an adaptive (online) 
linear approximation model and an adaptive bilinear approximation
model (for the Koopman system) updated by an iterative algorithm for
parameter estimation. In Section 4, we employ the developed linear and
bilinear Koopman system models to transform the original nonlinear
programming control problem to a linear programming problem.

\setcounter{equation}{0}
\section{Koopman System and Model Approximation}
\label{discrete}
In this section, we study the Koopman operator definitions for discrete-time
systems, the dynamic equations (Koopman equations) induced by Koopman
operation on observation functions of the state vector of a dynamic
system, the dynamic (functional) transformation using
observation functions as the state variables (Koopman variables) of a
new dynamic system (Koopman system), and the iterative (adaptive)
learning algorithms for linear or bilinear approximation of a
nonlinear Koopman system, to be used for feedback control design.

\subsection{Koopman Operator}
We start with a discrete-time nonlinear system without a control
input, described by
\beq
x(k+1) = f(x(k)),
\label{dsys}
\eeq
where $k=0,1,2,\ldots,$ are the discrete time instants, $x(k) \in R^n$
is the system state vector, and $f(x(k)) \in R^n$ represents the
nonlinear dynamic vector function of the system. The simplified notation 
$x(k+1) = f(x(k))$ is used to denote the discrete-time system:
$x(t_{k+1}) = f(x(t_k))$, $k=0,1,2,\ldots$, for a system state
vector $x(t)$ at the discrete-time instant $t = t_k$.

\subsubsection{Definition Based on $x(k+1) = f(x(k))$}
An observation of the state vector $x(k)$ can be made through
a scalar function $\psi_0(\cdot)$ of $x(k)$: $\psi_0(x(k)) \in R$ for
$x(k) \in R^n$, called an observable. An operation on or a 
transformation of the observable $\psi_0(x)$ can be made,
in terms of the dynamic function $f(x)$ of the system
(\ref{dsys}), by the Koopman operator originally proposed in
\cite{k31}, which may be described by the following definition, 
in terms of the dynamic system (\ref{dsys}).

\begin{defn} 
\label{dko}
{\rm \cite{km18}, \cite{mms20} The Koopman operator ${\cal K}$, applied
 on an observable $\psi_0(x) \in R$, for $x \in R^n$ from the
 system (\ref{dsys}) with a dynamic function $f(x) \in R^n$, is
 defined by the Koopman operation:
\beq
({\cal K} \psi_0)(x) = {\cal K} [\psi_0(x)] = \psi_0(f(x)).
\label{ko0}
\eeq}
\end{defn} 

Definition \ref{dko}, when applied to the system (\ref{dsys}), implies
(defines) the equation (identity):
\beq
{\cal K} [\psi_0(x(k))] = \psi_0(f(x(k))) = \psi_0(x(k+1)),
\label{koe}
\eeq
which shows three important features of the Koopman operator (operation): 

\begin{quote}
$\bullet$ it defines a system-related composite functional operation
  on $x(k)$, by ${\cal K} [\psi_0(x(k))] = \psi_0(f(x(k)))$, as $f(x(k))$
  is the system dynamic function; 

$\bullet$ it defines a time-advance operation, by ${\cal K}
  [\psi_0(x(k))] = \psi_0(x(k+1))$, as $x(k+1)$ is the time-advanced
  (future) value of the system state vector $x(k)$; and 

$\bullet$ it defines a new dynamic relationship, by $\psi_0(f(x(k))) =
  \psi_0(x(k+1))$, as it involves both time instants $k$ and $k+1$.
\end{quote}

Suppose that $N$ observations (observables) $\psi_i(x(k))$
($i=1,2,\ldots, N$) have been made for some chosen functions 
$\psi_i(\cdot)$ of certain interest.
 Then, under the Koopman operation (\ref{ko0}), we
have a set of Koopman equations:
\beq
{\cal K} [\psi_i(x(k))] = \psi_i(f(x(k))) =
\psi_i(x(k+1)),\;i=1,2,\ldots, N,
\label{Nobs}
\eeq
which all involve both time instants $k$ and $k+1$ and thus represent some
new dynamic relations or a dynamic system, and can be further
considered to derive the following additional (key) feature of the
Koopman operation:

\begin{quote}
$\bullet$ 
it defines, when applied to a set of system observation
  functions $\psi_i(x(k))$ ($i=1,2,\ldots, N$), a dynamic system transformation, by $z_i(k)
  \stackrel{\triangle} = \psi(x(k))$ and $\psi_i(f(x(k))) =
  \psi_i(x(k+1)),\;i=1,2,\ldots, N$, as $\psi_i(f(x(k))) =
  \psi_i(x(k+1)) = z_i(k+1)$ is to be represented in terms of
  $\psi_j(x(k)) = z_j(k),\;j=1,2,\ldots, N$, with some functions
  $\xi_i(\cdot, \cdot, \ldots, \cdot)$: 
  $\psi_i(f(x(k))) = \xi_i(\psi_1(x(k)), \psi_2(x(k)),\ldots,
  \psi_N(x(k)))$.
\end{quote}

There is a vast amount of literature on Koopman operator theory and
its applications to dynamic system analysis and control, e.g.,
\cite{lm13}, \cite{m15}, \cite{bbpk16}, \cite{adm17}, \cite{lkb18}, 
\cite{bzb19}, \cite{am19}, \cite{hhv20}, \cite{km20},
\cite{bsh21}, \cite{bfv21}, also addressing the case when the dynamic
system is under the influence of a control input $u$.

\subsubsection{Definition Based on $x(k+1) = f(x(k), u(k))$}
\label{ktcs}
We now consider the controlled nonlinear system
\beq
x(k+1) = f(x(k),u(k)),
\label{dsysc}
\eeq
where $k=0,1,2,\ldots,$ are the discrete time instants, $x(k) \in R^n$
is the system state vector, $u(k) \in R^m$ is the system control input
vector, and $f(x(k), u(k)) \in R^n$ is the nonlinear dynamic function 
of the system with the control input $u$. Our next goal is to develop a
dynamic (functional) transformation for (\ref{dsysc}), which is also applicable to
the system (\ref{dsys}), leading to a Koopman dynamic system. 

\mdp
Inspired by the definition of the Koopman operator ${\cal K}$ in
(\ref{ko0}) associated with a dynamic system: $x(k+1) = f(x(k))$,  
${\cal K} [\psi_0(x(k))] = \psi_0(f(x(k)))$, for a scalar function
$\psi_0(x)$ of $x \in R^n$, which is essentially the
operation (\ref{koe}): ${\cal K} [\psi_0(x(k))] = \psi_0(x(k+1))$, we may
define the Koopman operator for the case when there is a control input
in the dynamic system equation, analogously.

\begin{defn} 
\label{dkoc}
{\rm The Koopman operator ${\cal K}$, on a scalar function
  $\psi_0(x(k))$ of $x(k) \in R^n$ from a dynamic system: $x(k+1) =
  f(x(k), u(k))$, is defined by the operation:
\beq
{\cal K} [\psi_0(x(k))] = \psi_0(x(k+1)).
\label{koc}
\eeq}
\end{defn} 

Definition \ref{dkoc}, when applied to the system (\ref{dsysc}), implies
(defines) the equation (identity):
\beq
{\cal K} [\psi_0(x(k))] = \psi_0(x(k+1)) = \psi_0(f(x(k), u(k))),
\label{koec}
\eeq
which introduces a new dynamic relationship
\beq
\psi_0(x(k+1)) = \psi_0(f(x(k), u(k))).
\label{koec1}
\eeq
The question is how to further explore and ``exploit''
 such a new dynamic equation.

\bp
Suppose that $N$ observations (observables) 
$\psi_i(x(k))$ ($i=1,2,\ldots, N$) have
been made for some chosen functions 
$\psi_i(\cdot)$. Then, under the Koopman operation (\ref{koec}), we
also have a set of Koopman equations:
\beq
{\cal K} [\psi_i(x(k))] = \psi_i(x(k+1)) = \psi_i(f(x(k),
u(k))),\;i=1,2,\ldots, N,
\label{Nobsc}
\eeq
of which there are three possible cases of interest to consider: 

\begin{quote}
$\bullet$ the transformation of the original nonlinear system to
another nonlinear system;  

$\bullet$ the transformation of the original nonlinear system to a linear
system with approximation errors (a practical form); and 

$\bullet$ the transformation of the original nonlinear system to a
bilinear system with approximation errors (a practical and
potentially more useful form).
\end{quote}

\subsection{Functional Transformations}
The Koopman operator ${\cal K}$ defined by the Koopman operation (\ref{koc})
generates a new system (Koopman system) based on an induced
functional transformation (a dynamic transformation).

\subsubsection{Functional Transformation: with A Diffeomorphism}
For a set of observations 
$\psi_i(x(k))$ ($i=1,2,\ldots, N=n$), denoted as $z_i(k) =
\psi_i(x(k)) (i=1,2,\ldots, n)$ (called the Koopman variables), 
we introduce the vectors 
\beq
z(k) = [z_1(k), z_2(k), \cdots, z_n(k)]^T \in R^n
\eeq
\beq
\psi(x(k)) = [\psi_1(x(k)),\psi_2(x(k)),\ldots,\psi_n(x(k))]^T \in R^n.
\eeq

If the mapping $z = \psi(x) \in R^n$ for $x \in R^n$: 
\beq
z(k) \stackrel{\triangle} = \psi(x(k)) \in R^n,\;x(k) \in R^n,
\label{z(k)0}
\eeq
is a diffeomorphism \cite{i95}, that is, $\frac{\partial \psi(x)}{\partial x}$ is
nonsingular, so that the inverse $\phi(\cdot) = \psi^{-1}(\cdot)$ of
$\psi(\cdot)$ is well-defined: $x = \psi^{-1}(z)$, for $z =
\psi(x)$. In this case, the vector function
\beq
\psi(f(x(k), u(k))) = [\psi_1(f(x(k),u(k))), \psi_2(f(x(k),u(k))),
  \ldots, \psi_n(f(x(k),u(k)))]^T \in R^n
\eeq
can be expressed as
\beq
\psi(f(x(k),u(k))) = \xi(\psi(x(k)), u(k)),
\label{nnon2}
\eeq
for some nonlinear vector function $\xi(\psi, u) \in R^n$ of $\psi \in R^n$
and $u \in R^m$. To see (\ref{nnon2}), with $z = \psi(x)$ in $\xi(\psi(x), u)$
and $x = \psi^{-1}(z)$ in $f(x,u)$ for $\psi(f(x,u)) =
\xi(\psi(x),u)$, we have $\psi(f(\psi^{-1}(z),u)) = \xi(z, u)$, that
is, with $z$ and $u$ being the variables, the function 
$\xi(z, u)$ can be constructed by
\beq
\xi(z, u) 
\stackrel{\triangle} = \psi(f(\psi^{-1}(z), u))
\eeq
which may be called a functional transformation, as a well-defined
transformation, given that $\psi^{-1}(\cdot)$, $f(\cdot, \cdot)$ and
$\psi(\cdot)$ are both well-defined functions.

Then, with $z(k) = \psi(x(k))$ as the state vector, from
(\ref{Nobsc}) and (\ref{nnon2}), we have
\beq
z(k+1) = \xi(z(k), u(k)), 
\label{zsysc}
\eeq
which is a dynamic system equation and may be called a Koopman system,
that is, we have obtained a new dynamic system (\ref{zsysc}) derived
from the original system (\ref{dsysc}), with the Koopman operation
(\ref{koc}) and the coordinate transformation (\ref{z(k)0}): 
$z(k) = \psi(x(k)) \in R^n$.

In this formulation, the observations $\psi_i(x(k))$ ($i=1,2,\ldots,
n$) only explicitly depend on the system state vector $x(k)$, not
on the system input $u(k)$, which is necessary for a state
(coordinate) transformation. Some of $z_i(k) = \psi_i(x(k))$
($i=1,2,\ldots,n$) may
be taken as the new output signals: $w(k) = C z(k)$, for some matrix
$\in R^{p \times N}$, to form a new control system:
\beq
z(k+1) = \xi(z(k), u(k)),\;w(k) = C z(k).
\eeq 

\subsubsection{Functional Transformation: the General Case}. 
For $N$ observations $z_i(k) = \psi_i(x(k))$ ($i=1,2,\ldots, N$), we
form the vectors
\beq
z(k) = [z_1(k), z_2(k), \ldots, z_N(k)]^T \in R^N
\eeq
\beq
\psi(x(k)) = [\psi_1(x(k)), \psi_2(x(k)), \ldots, \psi_N(x(k))]^T \in
R^N,
\eeq
and have the vector equation (identity): 
\beq
z(k) = \psi(x(k))
\eeq
which, if considered as an equation to solve for $x(k)$ with a given
$z(k)$, has a natural solution $x(k)$ (as from the observation
identity: $z(k) = \psi(x(k))$) and may have more than one solutions
$x(k)$ for the same $z(k)$ to meet $z(k) = \psi(x(k))$.\footnote{It
  however should not be understood as: for a given $\psi(\cdot)$, 
$z = \psi(x)$ has a solution $x$ for any chosen $z$.} For example, for a given observation equation
$z(k) = \psi(x(k))$, there may be a minimum-norm solution $x(k)$ to
satisfy this equation, similar to a linear equation $A x =
y$. Therefore, there exists a vector function $\phi(\cdot)$ such that
\beq
x(k) = \phi(z(k))
\label{phi}
\eeq
for $x(k) \in R^n$ and $z(k) \in R^N$ satisfying $z(k) =
\psi(x(k))$. Then, with $z = \psi(x)$ and $u$ being the variables, the function
$\xi(z, u)$ can be constructed as
\beq
\xi(z, u) \stackrel{\triangle} = \psi(f(\phi(z), u)) = \psi(f(x),
u)|_{x=\phi(z)} 
\eeq
which is well-defined as $\phi(\cdot)$, $f(\cdot, \cdot)$ and
$\psi(\cdot)$ are well-defined functions, so that, in general,
 we can also arrive at the functional transformation (\ref{nnon2}):
\beq
\psi(f(x(k),u(k))) = \xi(\psi(x(k)), u(k)),
\label{nnon3}
\eeq
for some function $\xi(\cdot, \cdot)$, to derive
the Koopman system dynamic equation (\ref{zsysc}): 
\beq
z(k+1) = \xi(z(k), u(k)). 
\label{zsysc1} 
\eeq

\medskip
\begin{rem}
\rm
If for $N > n$, a subvector $\bar{z}(k) \in R^n$ of $z(k) \in R^N$, $\bar{z}(k)
= \bar{\psi}(x(k))$ (where $\bar{\psi}(\cdot) \in R^n$ is the
corresponding subvector of $\psi(\cdot) \in R^N$) is a diffeomorphism
between $\bar{z}$ and $x$ (that is, $\frac{\partial \bar{\psi}(x)}{\partial x}$
is nonsingular, for $\bar{z} = \bar{\psi}(x)$), then, besides the
formulation in the above remark, we also have: $x(k) = \bar{\psi}^{-1}(\bar{z}(k))$, so that, 
with $z = \psi(x)$ in $\xi(\psi(x), u)$
and $x = \bar{\psi}^{-1}(\bar{z})$ in $f(x,u)$ for $\psi(f(x,u)) =
\xi(\psi(x),u)$, we have $\psi(f(\bar{\psi}^{-1}(z),u)) = \xi(z, u)$, that
is, with $z$ and $u$ being the variables, the function 
$\xi(z, u)$ can be constructed and expressed as
\beq
\xi(z, u) = \psi(f(\bar{\psi}^{-1}(\bar{z}), u)) = 
\psi(f(\bar{\psi}^{-1}(z), u)),
\eeq
where the second equality is only a nominal expression in terms of $z$
in $\bar{\psi}^{-1}(\cdot)$. \hspace*{\fill} $\Box$
\end{rem}

\medskip
In summary, we have the following conclusion.

\begin{prop}
\label{propo}
For a nonlinear system $x(k+1) = f(x(k), u(k))$ ($x \in R^n$, $u \in
R^m$) and a set of
observations $z_i(k) = \psi_i(x(k))$ ($i=1,2,\ldots, N \geq 1$), the Koopman
operation (\ref{koc}): ${\cal K} [\psi_i(x(k))] =
\psi_i(x(k+1))$, is associated with a functional transformation
(\ref{nnon3}): $\psi(f(x,u)) = \xi(\psi(x), u)$ with $\psi
= [\psi_1, \ldots, \psi_N]^T \in R^N$, and 
forms a dynamic system (\ref{zsysc1}): $z(k+1) = \xi(z(k), u(k))$. 
\end{prop}

\medskip
The Koopman operation generated system $z(k+1) = \xi(z(k), u(k))$ may
be called a Koopman dynamic system. In the Koopman system research, a
lot of effort has been applied to the task of approximating the Koopman
equations (\ref{Nobsc}): 
$\psi(x(k+1)) = \psi(f(x(k), u(k)))$, by a linear system equation: 
$\psi(x(k+1)) = A \psi(x(k)) + B u(k)$, i.e., $z(k+1) = A z(k) + B
u(k)$, which is a linear Koopman system such that 
$\xi(z(k), u(k)) = A z(k) + B u(k)$.

The equation $\psi(x(k+1)) = \psi(f(x(k), u(k)))$, a dynamic relation as
it involves both time instants $k$ and $k+1$, is a dynamic system
equation only when $\psi(f(x(k), u(k))) = \xi(\psi(x(k)), u(k))$ 
holds (so that $\psi(x(k+1)) = \xi(\psi(x(k)), u(k))$, or, 
$z(k+1) = \xi(z(k), u(k))$), which has been
formally established by Proposition \ref{propo} and is the foundation
for a linear Koopman system (and any other bilinear or multilinear
Koopman systems). 

Effort has also been applied to the task of selecting the
observation functions $\psi_i(x(k))$ such that the system function
$\xi(z(k), u(k))$ is naturally linear: $\xi(z(k), u(k)) = z(k+1) = A
z(k) + B u(k)$. 

\medskip
In this paper, we will focus our main attention on the first task:
approximation of a nonlinear Koopman system by either a linear system
model: $z(k+1) = A z(k) + B u(k) + \eta(k)$, or a bilinear system
model: $z(k+1) = A z(k) + B u(k) + D(x(k), u(k)) + \eta(k)$, where
$\eta(k)$ is the modeling error which is inevitable for many practical
problems, and $D(x(k), u(k))$ is a bilinear function of $x(k)$ and
$u(k)$, to be defined to capture more dynamics (and reduce the modeling
error $\eta(k)$) of the nonlinear Koopman
system: $z(k+1) = \xi(z(k), u(k))$.

\subsection{Linear Approximation of A Koopman System}
For a set of $N \geq n$ observations $\psi_i(x(k))$ ($i=1,2,\ldots,
N$), from the Koopman operation (\ref{koc}) and functional transformation
(\ref{nnon3}), the Koopman system dynamic functions 
\beq
{\cal K} [\psi_i(x(k))] = \psi_i(x(k+1)) = \psi_i(f(x(k), u(k))) = 
\xi_i(\psi(x(k)), u(k))
\eeq
may be expressed as
\beq
\xi_i(\psi(x(k)), u(k)) = a_i^T \psi(x(k)) + b_i^T u(k) +
\eta_i(k),\;i=1,2,\ldots, N,
\label{Nle2}
\eeq
for some parameter vectors $a_i \in R^N$ and $b_i \in R^m$,
$i=1,2,\ldots, N$, and 
\beq
\psi(x(k)) = [\psi_1(x(k)), \psi_2(x(k)), \ldots, \psi_N(x(k))]^T \in
R^N,
\eeq
where $\eta_i(k)$ is the approximation error of $\psi_i(f(x(k), u(k)))$
by $a_i^T \psi(x(k))) + b_i^T u(k)$. With $z_i(k) = \psi_i(x(k))$ and 
$z(k) = [z_1(k), z_2(k), \ldots, z_N(k)]^T \in R^N$, we have a linear model
\beq
z(k+1) = A\, z(k) + B \,u(k) + \eta(k),
\label{nsys2}
\eeq
with an approximation error vector $\eta(k)$,
where 
\beq
A = \left[ \begin{array}{c}
a_1^T\\
\vdots\\
a_N^T
\end{array}
\right] \in R^{N \times N},\;
B = \left[ \begin{array}{c}
b_1^T\\
\vdots\\
b_N^T
\end{array}
\right] \in R^{N \times m}
\label{AB}
\eeq
\beq
\eta(k) = [\eta_1(k), \eta_2(k),\ldots, \eta_N(k)]^T \in R^N.
\eeq
Such a linear model has been commonly used in the Koopman
system literature, for approximation of the Koopman equations 
(\ref{Nobsc}): $\psi_i(x(k+1)) = \psi_i(f(x(k),
u(k))),\;i=1,2,\ldots, N$.

\medskip
In (\ref{nsys2}), the parameters $A$ and $B$ are expected to be such
that the error $\eta(k)$ is minimized. This motivates several methods to
calculate $A$ and $B$. 

\mdp
{\bf Method I}: 
One method is to find $A$ and $B$ to minimize the instantaneous error sum
\beq
J_1 = \sum_{i=1}^{N} \eta_i^2(k) = 
\eta^T(k) \eta(k) = (z(k+1) - A z(k) - B u(k))^T (z(k+1) - A z(k) - B u(k)).
\eeq

\mdp
{\bf Method II}: 
Another method \cite{km18} is to find $A$ and $B$ to minimize the
accumulative error
sum over a signal measurement interval $k = k_0, k_0+1, \ldots, k_0 + k_1$:
\beq
J_2 = \sum_{k=k_0}^{k_0+k_1} \sum_{i=1}^{N} \eta_i^2(k) = 
\sum_{k=k_0}^{k_0+k_1} (z(k+1) - A z(k) - B u(k))^T (z(k+1) - A z(k) -
B u(k)).
\label{J2}
\eeq

The solution $[A, B]$ minimizing $J_2$ satisfies
\beq
[A, B]\; \sum_{k=k_0}^{k_0+k_1} 
\left[ \begin{array}{c}
z(k)\\
u(k)
\end{array}
\right] 
\left[ \begin{array}{c}
z(k)\\
u(k)
\end{array}
\right]^T = 
\sum_{k=k_0}^{k_0+k_1} 
z(k+1)
\left[ \begin{array}{c}
z(k)\\
u(k)
\end{array}
\right]^T.
\label{soleq}
\eeq

The solution $[A, B]$ meeting (\ref{soleq}) may not be unique and
needs to be re-calculated when $k_1$ increases, that is, when new
measurements (data) are collected, with increasing computations.
 
\mdp
{\bf Method III}: 
The third method is to find the iteratively updated (online) estimates
$\hat{a}_i(k)$ and $\hat{b}_i(k)$ of $a_i$ and $b_i$, as $k$ goes,
from the incoming measurements (data) of $z(k+1)$, $z(k)$ and
$u(k)$, with fixed computations. An iterative (adaptive) algorithm
will be given in this paper, 
which will be extensively simulated for identification and optimal
control for the system of multiple robots with nonlinear utility
functions to be selected as a group of Koopman variables $z_i(k)$.

\medskip
\begin{rem}
\rm
The ideal case of (\ref{nsys2}) is with $\eta(k) \equiv 0$: 
\beq
z(k+1) = A\, z(k) + B \,u(k),
\label{ilsm}
\eeq
as the (exact) representation of the
Koopman system (\ref{zsysc1}): $z(k+1) = \xi(z(k), u(k))$, under the
transformation $\xi(z(k), u(k)) = \psi(f(x(k), u(k))) = \psi(x(k+1)) = 
z(k+1)$. This can happen when the observations $\psi_i(x(k))$ (as the
components of $\psi(x(k))$) and the system function $f(x(k), u(k))$
(of the original system $x(k+1) =  f(x(k), u(k))$) are correlated in
a perfect matching, to make $\xi(z(k), u(k)) = A\, z(k) + B \,u(k)$. 
In other words, for a given $f(x(k), u(k))$, the observations 
$\psi_i(x(k)) (i=1,2,\ldots, N)$ should be carefully chosen, if
possible, to make an ideal linear system model (\ref{ilsm}). This has
been a dream and yet hard task for the Koopman system research,
only fulfilled for some simple example systems. 
\hspace*{\fill} $\Box$
\end{rem}

\medskip
While the smallness of the approximation error $\eta(k)$ in
(\ref{nsys2}) is crucial for many applications, the linear model
approximation structure in (\ref{nsys2}) can have certain limitations
in some applications. In other words, the approximation of 
$z(k+1) =  \xi(z(k), u(k))$ by a linear model
\beq
\xi(z(k), u(k)) = A\, z(k) + B \,u(k) + \eta(k) = [A, B]
 \left[ \begin{array}{c}
z(k)\\
u(k)
\end{array}
\right] + \eta(k),
\label{lsm}
\eeq
or by a bilinear model
\beq
\xi(z(k), u(k)) = A\, z(k) + B \,u(k) + 
\left[ \begin{array}{c}
[z^T(k), u^T(k)] D_1\\ 

\vdots \\

[z^T(k), u^T(k)] D_N 
\end{array}
\right] 
\left[ \begin{array}{c}
z(k)\\
u(k)
\end{array}
\right]
+ \eta_b(k),
\label{blsm}
\eeq
for some additional parameter matrices $D_i$ ($i=1,2,\ldots, N$), 
can make a big difference, because 

\begin{quote}
$\bullet$ the model errors $\eta(k)$ and $\eta_b(k)$ can be different
  as obvious; and 

$\bullet$ even if the model errors $\eta(k)$ and $\eta_b(k)$ are not
  different under certain conditions, the model structures are
  different, e.g., the 
  distributions of $\xi(z(k), u(k))$ over $z(k)$ and $u(k)$ are
  different for different models, which can have some significant
  effect on control system performance, when the models are used for
  feedback control design.
\end{quote}

Thus, it is desirable to study a bilinear model approximation of a
Koopman system.

\subsection{Bilinear Approximation of A Koopman System}
Considering a vector function $f(x) = [f_1(x), f_2(x), \ldots,
  f_m(x)]^T \in R^m$ of $x = [x_1, x_2, \ldots, x_n]^T \in R^n$
(which, here, is a generic notation, not related to a system $x(k+1) = f(x(k),
u(k))$), and applying
Taylor's theorem at a chosen point $x_0 \in R^n$ of $x$, we express
\beq
f_i(x) = f_i(x_{0}) + (\nabla f_i(x_0))^T (x - x_0) + \frac{1}{2} (x -
x_0)^T H_{f_i}(x_0) (x - x_0) + \mbox{higher-order terms},
\label{fi}
\eeq
for $i=1,2,\ldots, n$, where 
\beq
\nabla f_i(x_0) = \frac{\partial f_i}{\partial x}|_{x = x_0} = 
[\frac{\partial f_i}{\partial x_1}|_{x = x_0}, 
\frac{\partial f_i}{\partial x_2}|_{x = x_0}, \ldots, 
\frac{\partial f_i}{\partial x_n}|_{x = x_0}]^T \in R^{n}
\label{nablafi(x0)}
\eeq
\beq
H_i = H_{f_i}(x_0) = [h_{f_i,pq}]_{p,q=1,2,\ldots,n} = [\frac{\partial^2 f_i}{\partial x_p \partial
  x_q}|_{x= x_0}] 
= \left[\begin{array}{ccc}
\frac{\partial^2 f_i}{\partial x_1 \partial x_1} & \cdots & 
\frac{\partial^2 f_i}{\partial x_1 \partial x_n}\\
& \vdots & \\
\frac{\partial^2 f_i}{\partial x_n \partial x_1} & \cdots & \frac{\partial^2 f_i}{\partial x_n \partial x_n}
\end{array}
\right]_{x = x_0} \in R^{n \times n},
\label{Hi0}
\eeq
which is the (generically symmetric) Hessian matrix of $f_{i}(x)$ at
$x_0$, that is, $H_i = H_{f_i} = H_{f_i}^T = H_i^T$.

We then apply this approximation expression to the Koopman system
(\ref{zsysc1}): 
\beq
z_i(k+1) = \xi_i(z(k), u(k)),\;i=1,2,\ldots, N,
\label{zi}
\eeq
for $f_i(x) = \xi_i(z(k), u(k))$ and $x = [z^T, u^T]^T$ ($i=1,2,\ldots, N$).

With $f_i(x_{0}) = 0$, $(\nabla f_i(x_0))^T = [A, B]$, 
$x_0 = 0$ (a zero vector) and $D_i = \frac{1}{2} H_{f_i}$, we can see
that (\ref{fi}) and (\ref{zi}) lead to the bilinear model
(\ref{blsm}).

\medskip
As a comparison, the linear model approximation (\ref{lsm}) is 
based on the approximation
\beq
f_i(x) \approx (\nabla f_i(x_0))^T (x - x_0),\;x_0 = 0_{1 \times n},
\eeq
which is a special case of the bilinear expansion (\ref{fi}),
and, when applied to the robot system with nonlinear utility
functions, may not provide enough capacity to solve the control
problem of driving the robots to their desired positions and
velocities. 

The bilinear expansion (\ref{fi}) will be studied in Section 3 for the 
multi-robot system.

\setcounter{equation}{0}
\section{Koopman Models of A Multi-Robot System}
\label{robot}
In this section, we address the problem of developing a linear or
bilinear model for a $5$-robot system with nonlinear utility functions
for optimal control. Our plan is to choose a set of Koopman variables
(consisting of the set of robot system state variables and the set of
the utility functions plus possibly an additional set of state
variables), to form a Koopman system for the multi-robot system, and
to develop iterative (online) approximations of the Koopman system,
using a linear model structure and a bilinear model structure. We will
then formulate the optimal control problem based on the linear or
bilinear model of the utility functions, to approximate the original
optimal control of maximizing the nonlinear utility functions.

In Section 3.1, we study the $5$-robot system model combined with the
nonlinear utility functions introduced in the Ford report 
\cite{wetal20} for optimal control of the robot system. In Section
3.2, we introduce the Koopman variables to form a Koopman system for
the combined robot system. In Section 3.3, we develop an iterative
algorithm for estimating the parameters of a linear Koopman system
model. In Section 3.4, we develop an iterative algorithm for
estimating the parameters of a bilinear Koopman system model. In both
sections, we also describe a simulation study plan. In
Section 3.5, we formulate the optimal control based on a linear or
bilinear Koopman system model.

\subsection{Robot System and Utility Functions}
We now present the multi-robot system model with nonlinear utility
functions (to be maximized to generate the control signal to drive 
the robots to some desired locations).

\bp
{\bf Robot system model}. A 5-robot system model is described as
\bea
v_{i}(k+1) \ts = \ts v_{i}(k) + a_{i}(k)\, \Delta t \nn\\
r_{i}(k+1) \ts = \ts r_{i}(k) + v_{i}(k)\, \Delta t + 0.5\,
a_{i}(k)\,(\Delta t)^2,\;i=1,2,\ldots, 5,
\label{robots3}
\eea
where $r_{i}(k) = (x_{i}(k), y_{i}(k))$ are the $i$th robot's
positions, $v_{i}(k) = (v_{i}^x(k), v_{i}^y(k))$ are the velocities, 
$a_{i}(k) = (a_{i}^x(k), a_{i}^y(k))$ are the acceleration vectors as
the control vector signals to the robots, and $\Delta t$ is a fixed
number representing a uniform sampling interval. 

With the total state vector defined by
\beq
X(k) = [x_{1}(k), y_{1}(k), \ldots, x_{5}(k), y_{5}(k), v_{1}^x(k),
  v_{1}^y(k), \ldots, v_{5}^x(k), v_{5}^y(k)]^T \in R^{20},
\label{Xk}
\eeq
we can express (\ref{robots3}) as
\beq
X(k+1) = A X(k) + B U(k),
\label{Xequation3}
\eeq
where the control vector $U(k)$ is
\beq
U(k) = [a_{1}^x(k), a_{1}^y(k), a_{2}^x(k), a_{2}^y(k), \ldots,
  a_{5}^x(k), a_{5}^y(k)]^T \in R^{10},
\eeq
and $A \in R^{20 \times 20}$ and $B \in R^{20 \times 10}$ are some
matrices as the controlled system matrices which can be directly
obtained from the system model (\ref{robots3}).

\bp
{\bf Utility functions}. The utility
functions of \cite{wetal20} have the form
\beq
u_{i}(k) = \sum_{j=1}^6 \omega_i^{(j)} \phi_{i}^{(j)}(X(k)),
\label{utility2}
\eeq
where $X(k)$ in (\ref{Xk}) depends on $s_{l}(k) = \{r_{l}(k),
v_{l}(k)\}$ ($l=1,2,\ldots, 5$), $\omega_i^{(j)}$ are 
constants, and the details of the nonlinear functions
 $\phi_{i}^{(j)}(X(k))$ are given in
\cite{wetal20} (and also in this paper's companion paper \cite{zth22}: 
``Koopman system approximation based optimal control of multiple
  robots --- Part II: Simulations and Evaluations), 
in terms of $s_{l}(k) = \{r_{l}(k), v_{l}(k)\}$ ($l=1,2,\ldots, 5$). The
optimal control problem is to find the control signal $U(k)$ to
maximize the utility functions $u_i(k+1) (i=1,2, \ldots, 5)$.

Inspired by \cite{km18}, a set of Koopman variables are defined as
\beq
z_{i,j}(k) = \psi_{i,j}(X(k)) =  \phi_{i}^{(j)}(X(k)),
\label{zij}
\eeq
for $i=1,2,\ldots, 5,\;j=1,2,\ldots, 6$, and the corresponding
Koopman equations are
\beq
\psi_{i,j}(X(k+1)) =  \psi_{i,j}(A X(k) + B U(k)) = 
\phi_{i}^{(j)}(A X(k) + B U(k)).
\eeq

A main goal of our study is to express 
$\phi_{i}^{(j)}(A X(k) + B U(k))$ in terms of 
$\psi_{i,j}(X(k))$ (plus some other variables including $X(k)$) and $U(k)$, so that $z_{i,j}(k+1) = \psi_{i,j}(X(k+1))$ can
be expressed in terms of $z_{i,j}(k) = \psi_{i,j}(X(k))$ and other
variables and signals. This will transform the nonlinear utility
functions $\phi_{i}^{(j)}(X(k))$ in (\ref{utility2}) to a set of
linear functions $z_{i,j}(k)$.

\subsection{Koopman System Construction}
A key step in applying the Koopman system techniques to system
modeling and feedback control is to select the desired components
$z_i(k)$ of $z(k)$, the Koopman variables.

\subsubsection{Koopman System with Natural Variables}
There are two sets of natural system variables: one set from the robot
system state variables, and one set from the utility functions, and
both can be used as the Koopman system variables.

\bp
{\bf Robot system state variables}. In our study, we also take the original
system state vector $X(k)$ in (\ref{Xk}) as a part of the Koopman
state vector $z(k)$:
\beq
z_i(k) = \psi_i(X(k)) = X_i(k),\;i=1,2,\ldots, 20,
\label{zig1}
\eeq
where $X_i(k)$ ($1,2,\ldots, 20$) are the components of $X(k)
\stackrel{\triangle} = [X_1(k), X_2(k), \ldots, X_{20}(k)]^T$.

\medskip
Then, the Koopman
equations (\ref{Nobsc}): $\psi_i(f(x(k), u(k))) = \psi_i(x(k+1)) =
z_i(k+1)$ for $x(k+1) = f(x(k), u(k))$ (i.e., $X(k+1) = A X(k) +
B U(k)$) become
\[
\psi_i(A X(k) + B U(k)) = \psi_i(X(k+1))\]
\beq
 \stackrel{\psi_i(X(k)) =
  X_i(k)} \Longrightarrow (A X(k) + B U(k))_i = X_i(k+1) = z_i(k+1),
\eeq
where $(A X(k) + B U(k))_i$ denotes the $i$th component of $A X(k) + B
U(k)$, or in the vector form, 
\beq
A X(k) + B U(k) = X(k+1) \Leftrightarrow 
A z_{(1)}(k) + B U(k) = z_{(1)}(k+1),
\label{zisys}
\eeq
where 
\beq
z_{(1)}(k) = [z_1(k), \ldots, z_{20}(k)]^T = X(k) \in R^{20},
\label{z(1)}
\eeq
that is, the Koopman subsystem (\ref{zisys}) for $z_{(1)}(k) = [z_1(k), \ldots,
  z_{20}(k)]^T$ is the original robot system (\ref{Xequation3}) which
is already linear in this case without any transformation.

\bp
{\bf Utility functions as Koopman variables}. Then, we formally include the
previously defined Koopman variables: the $5$ sets of the utility
function components $\phi_{i}^{(j)}(X(k))$, each set for one robot
 with $6$ components, for $i=1,2,\ldots, 5,\;j=1,2,\ldots, 6$:
\bea
\ts \ts z_{20+6(i-1)+j}(k) = \psi_{20+6(i-1)+j}(X(k)) \nn\\
\ts = \ts z_{i,j}(k) =
\psi_{i,j}(X(k)) = \phi_{i}^{(j)}(X(k)),\;i=1,2,\ldots, 5,\;j=1,2,\ldots, 6, 
\label{zig2}
\eea
(from $z_{21}(k)$ to $z_{50}(k)$),
 for $\phi_{i}^{(j)}(X(k))$ in the utility function (\ref{utility2}),
 whose details are shown in \cite{zth22}, 
to transform the utility functions to a linear form:
\beq
u_{i}(k) = \sum_{j=1}^6 \omega_i^{(j)} z_{20+6(i-1)+j}(k),
\label{lf}
\eeq
in terms of the Koopman system variables $z_{20+6(i-1)+j}(k)$,
$i=1,2,\ldots, 5,\;j=1,2,\ldots, 6$. 

\bp
For the utility function components based definitions of Koopman
variables $z_{20+6(i-1)+j}(k)$ introduced in (\ref{zig2}): 
$z_{20+6(i-1)+j}(k) = \psi_{20+6(i-1)+j}(X(k)) =
\phi_{i}^{(j)}(X(k))$, the Koopman equations (\ref{Nobsc}):
$\psi_i(f(x(k), u(k))) = \psi_i(x(k+1)) = z_i(k+1)$ for $x(k+1) =
f(x(k), u(k))$, applied to $X(k+1) = A X(k) + B U(k)$, become
\[
\psi_l(A X(k) + B U(k)) = \psi_l(X(k+1)) \]
\beq
\stackrel{z_i(k) = \psi_l(X(k)) =
  \phi_{i}^{(j)}(X(k))} \Longrightarrow \phi_{i}^{(j)}(A X(k) + B U(k)) =
 \phi_{i}^{(j)}(X(k+1)) = z_l(k+1),
\eeq
for $l= 20+6(i-1)+j+1 = 21, 22, \ldots, 50$ (with $i=1,2,\ldots,
5,\;j=1,2,\ldots, 6$). 

Similar to $z_{(1)}(k)$ in (\ref{z(1)}), we can define 
\beq
z_{(2)}(k) = [z_{21}(k), \ldots, z_{50}(k)]^T = \psi_{(2)}(X(k)) =
[\psi_{21}(X(k)), \ldots, \psi_{50}(X(k))]^T \in R^{30},
\label{z(2)}
\eeq
as the second group of Koopman variables. As shown in Section
\ref{ktcs}, the Koopman equations (\ref{Nobsc}):
$\psi_i(f(x(k), u(k))) = \psi_i(x(k+1)) = z_i(k+1)$ for $x(k+1) =
f(x(k), u(k))$ (i.e., $X(k+1) = A X(k) + B U(k)$)
 can be expressed as (\ref{zsysc}): $z(k+1) = \xi(z(k),
u(k))$, for some vector function $\xi(\cdot, \cdot)$. This implies,
when applied to $\phi_{i}^{(j)}(A X(k) + B U(k)) =
\phi_{i}^{(j)}(X(k+1))$, that
\beq
z_{(2)}(k+1) = \xi_{(2)}(z_{(1)}(k), z_{(2)}(k), U(k))
\label{xi(2)}
\eeq
for some vector function $\xi_{(2)}(\cdot, \cdot, \cdot) \in R^{30}$,
where $z_{(2)}(k) = \psi_{(2)}(X(k))$, and 
\beq
z_{(2)}(k+1) = \psi_{(2)}(X(k+1)) = \psi_{(2)}(A X(k) + B U(k)).
\label{xi(2)1}
\eeq

\subsubsection{Koopman System with Possible Additional State Variables} 
While the use of the robot system state variables $z_{(1)}(k) = X(k)$
as a part of the Koopman system state vector $z(k)$ can help expanding
the basis of Koopman variables beyond those from $z_{(2)}(k)$,
additional Koopman system state variables $z_i(k) = \psi_i(X(k))$,
$k=51, 52, \ldots,
N$, if properly chosen, can help further to increase the accuracy of
the linear or bilinear model approximation of the $5$-robot system
with nonlinear utility functions. Neural network functions can be good
candidates for such Koopman variables (observations) $z_i(k) =
\psi_i(X(k))$ \cite{lkb18}, \cite{s06}.

The current choice of $z_{(1)}(k)$
and $z_{(2)}(k)$ is to build a basic framework: a dynamic system for $z(k) =
[z_{(1)}^T(k), z_{(2)}^T(k)]^T$ in (\ref{zisys}) and (\ref{xi(2)}),
with the utility functions in (\ref{utility2}) expressed as
(\ref{lf}), and eventually, as the total utility function
\beq
u(k) = \sum_{i=1}^6 u_i(k) = \sum_{i=1}^6 \omega_i^{T} z_{(2)i}(k) =
\omega^T z_{(2)}(k),
\eeq
for $z_{(2)}(k) = [(z_{(2)1}(k))^T, \ldots, (z_{(2)5}(k))^T]^T \in
R^{30}$ for $z_{(2)i}(k) \in R^6$, $\omega_i = [\omega_i^{(1)}, \ldots,
  \omega_i^{(6)}]^T \in R^6$, $i = 1,\cdots,5$, and $\omega =
[\omega_1^T, \ldots, \omega_5^T]^T \in R^{30}$. In this way, the
nonlinear utility functions $u_i(k)$ are linear in the new system
variables $z_{(2)}(k)$. 

Recall that the control objective is to find $U(k)$ to maximize
$u(k+1)$ subject to some constraints on $U(k)$, originally, for 
$z_{(2)}(k+1)$ from the nonlinear model (\ref{xi(2)1}) (nonlinear in
the control vector $U(k)$), and, based on
the study of this paper (with a linear or bilinear model
approximation of the Koopman system), for $z_{(2)}(k+1)$ from either a
linear or bilinear model, to make $u(k+1)$ linear in or affine in
$U(k)$, simplifying the optimal control solution procedure.

With the possible addition of more Koopman variables $z_i(k)$, $k=51,
52, \ldots, N$, from some additional observations $z_i(k) =
\psi_i(x(k)) = \psi_i(X(k))$, to form $z_{(3)} = [z_{51}, \ldots,
  z_{N}]^T$, for some $N > 50$, the total Koopman system state vector
is $z(k) = [z_{(1)}^T(k), z_{(2)}^T(k), z_{(3)}^T(k)]^T$, and while the
utility xfunction is still the same $u(k) = \omega^T z_{(2)}(k)$. In
this case, there are two subsystems to be approximated: 
\beq
z_{(2)}(k+1) = \xi_{(2)}(z_{(1)}(k), z_{(2)}(k), z_{(3)}(k), U(k)),
\eeq
\beq
z_{(3)}(k+1) = \xi_{(3)}(z_{(1)}(k), z_{(2)}(k), z_{(3)}(k), U(k)),
\eeq
where $\xi_{(i)}(\cdot, \cdot, \cdot, \cdot)$, $i=2,3$, are some
vector functions which are ensured to exist by Proposition
\ref{propo}, for $z_{(3)}(k)$ being a vector of additional
observations: $z_{(3)}(k) = z_{(3)}(X(k))$. 

Hence, the basic solution procedure will be the same, if some
additional koopman variables $z_i(k) = z_i(X(k))$, $k=51, 52, \ldots,
N$, in $z_{(3)}(k) = [z_{51}(k), z_{52}(k), \ldots, z_N(k)]^T$, 
are to be included. The expectation is that the addition of 
$z_{(3)}(k)$ makes the approximation of $z_{(2)}(k)$ closer to
$z_{(2)}(k)$, in the sense that a linear programming control design
based on approximation of $z_{(2)}(k)$ solves the original nonlinear
programming control problem for $z_{(2)}(k) = \Psi_{(2)}(X(k))$ being
nonlinear more accurately than that without the addition of 
$z_{(3)}(k)$.

\subsection{Iterative Algorithm for Linear Model Identification}
\label{linearam}
In this subsection, we develop an iterative algorithm for estimating
the parameters of a linear approximation model of the robot system
with nonlinear utility functions, for the Koopman system model
(\ref{xi(2)}), based on the Koopman variables $z_{(1)}(k)$ and
$z_{(2)}(k)$.

\subsubsection{Linear Model Structure}
We form the augmented Koopman vector vector 
\beq
z(k) = [(z_{(1)}(k))^T, (z_{(2)}(k))^T]^T \in R^{50}
\label{z(k)}
\eeq
for $z_{(1)}(k) \in R^{20}$ in (\ref{z(1)}) and $z_{(2)}(k) \in
R^{30}$ in (\ref{z(2)}). Our interest is to find some matrices $A_z \in
R^{50 \times 50}$ and $B_z \in R^{50 \times 10}$ such that 
\beq
z(k+1) = A_z \,z(k) + B_z\,U(k) + \eta_z(k),
\label{zs1}
\eeq
for some minimized approximation error vector $\eta_z(k) \in R^{50}$. From
(\ref{zisys}), we can see that $A_z$, $B_z$ and $\eta_z(k)$ have the
following structures:
\beq
A_z =  \left[ \begin{array}{cc}
A & 0_{20 \times 30}\\
A_{21} & A_{22}
\end{array}
\right],\;
B_z =  \left[ \begin{array}{c}
B\\
B_2
\end{array}
\right],\;
\eta_z(k) =  \left[ \begin{array}{c}
0_{20 \times 0}\\
\eta_{(2)}(k)
\end{array}
\right],
\label{AzBz}
\eeq
where $A_{21} \in R^{30 \times 20}$, $A_{22} \in R^{30 \times 30}$ and
  $B_2 \in R^{30 \times 10}$ are parameter matrices to be calculated
  to minimize the error vector $\eta_{(2)}(k) \in R^{30}$.

Three methods have been given in Section \ref{ktcs} for calculating or
estimating such matrices $A_z$ and $B_z$. The second method minimizing
the cost function $J_2$ given in (\ref{J2}) can be used, taking into
account the special 
form of $A_z$ and $B_z$ in (\ref{AzBz}) with $A$ and $B$ known, given
that the signals $z(k+1)$ and $z(k)$ are measured signals: 
$z_{(1)}(K) = X(k)$ and $z_{(2)}(k)$ with components
$\phi_{i}^{(j)}(X(k))$. The specific algorithm for obtaining $A_{21}$,
$A_{22}$ and $B_2$ can be derived.

The cost function $J_2$ given in (\ref{J2}) is
\beq
J_2 = \sum_{k=k_0}^{k_0+k_1} \eta_z^T(k) \eta_z(k) = 
\sum_{k=k_0}^{k_0+k_1} (z(k+1) - A_z z(k) - B_z U(k))^T (z(k+1) - A_z z(k) -
B_z U(k)), 
\eeq
as the sum of $\eta_z(k)$ over $[k_0, k_0+k_1]$. 
From (\ref{AzBz}), we have
\beq
z_{(1)}(k+1) = A z_{(1)}(k) + B U(k)
\label{z(1)(k+1)0}
\eeq
\beq
z_{(2)}(k+1) = A_{21} z_{(1)}(k) + A_{22} z_{(2)}(k) + B_2  U(k) +
\eta_{(2)}(k),
\label{z(2)(k+1)}
\eeq
so that the above cost function $J_2$ becomes
\bea
J_2 \ts = \ts \sum_{k=k_0}^{k_0+k_1} \eta_{(2)}^T(k) \eta_{(2)}(k) = 
\sum_{k=k_0}^{k_0+k_1} (z_{(2)}(k+1) - A_{21} z_{(1)}(k) - A_{22}
z_{(2)}(k) - B_2 U(k))^T \nn\\
\ts \ts \cdot (z_{(2)}(k+1) - A_{21} z_{(1)}(k) - A_{22}
z_{(2)}(k) - B_2 U(k)).
\label{J21}
\eea
One can find the matrices $A_{21}$, $A_{22}$ and $B_2$ to minimize
$J_2$, using a batch data set of $z(k)$ and $U(k)$ over an interval
$[k_0, k_0+k_1]$. To make use of the incoming new data, one can
re-solve the minimization problem with increased $k_1$ (and
$k_0$). Such an algorithm for obtaining the matrices $A_{21}$,
$A_{22}$ and $B_2$ is a batch-data algorithm.

It would be useful to derive an iterative algorithm which can update
the values of $A_{21}$, $A_{22}$ and $B_2$, using the incoming new
data and the previous values of $A_{21}$, $A_{22}$ and $B_2$. Such an
algorithm has certain online learning properties desirable for
applications.

\subsubsection{Iterative Algorithm for the 5-Robot System}
\label{iterative}
For the Koopman state vector $z(k)$ defined in (\ref{z(k)}), we want
to the matrices $A_z$ and $B_z$ in (\ref{zs1}): 
$z(k+1) = A_z \,z(k) + B_z\,U(k) + \eta_z(k)$, to minimize the error 
$\eta_z(k)$ in some sense. From the structures of $A_z$ and $B_z$ in 
(\ref{AzBz}), we are led to the linear model (\ref{z(2)(k+1)}) with 
$\eta_{(2)}(k)$:
\beq
z_{(2)}(k+1) = A_{21} z_{(1)}(k) + A_{22} z_{(2)}(k) + B_2  U(k) +
\eta_{(2)}(k),
\label{z(2)(k+1)2}
\eeq
and to find the parameter matrices 
$\hat{A}_{21} \in R^{30 \times 20}$, $\hat{A}_{22} \in R^{30 \times 30}$ and
  $\hat{B}_2 \in R^{30 \times 10}$, to minimize 
$\eta_{(2)}(k) \in R^{30}$ in the sense that some accumulative sum $J$
of the errors 
\beq
z_{(2)}(\tau+1) - (\hat{A}_{21} z_{(1)}(\tau) + \hat{A}_{22} z_{(2)}(\tau) + \hat{B}_2
U(\tau))
\eeq
is minimized over the time interval: $k_0, k_0+1, \ldots, k-1$.

To present an iterative algorithm for $\hat{A}_{21} \in R^{30 \times
  20}$, $\hat{A}_{22} \in R^{30 \times 30}$ and $\hat{B}_2 \in R^{30
  \times 10}$, we let
\beq
y(k) = z_{(2)}(k+1) \in R^{30}
\eeq
\beq
\zeta(k) = [z_{(1)}^T(k), z_{(2)}^T(k), U^T(k)]^T \in R^{60}
\label{zeta}
\eeq
\beq
\Theta^T = [A_{21}, A_{22}, B_2] \in R^{30 \times 60}
\label{Theta}
\eeq
and rewrite (\ref{z(2)(k+1)2}) as
\beq
y(k) = \Theta^T \zeta(k) + \eta_{(2)}(k),
\label{y(k)}
\eeq
where
\beq
\eta_{(2)}(k) \stackrel{\triangle} = [\eta_{2,1}(k), \eta_{2,2}(k),
  \ldots, \eta_{2,30}(k)]^T \in R^{30}.
\eeq

The cost function $J_3$ to be minimized is chosen as
\bea
J_3 = J_3(\hat{\Theta}) \ts = \ts \frac{1}{2} \sum_{\tau = k_0}^{k-1} 
\frac{1}{\rho} (\hat{\Theta}^{T}(k) \zeta(\tau) - y(\tau))^{T}
(\hat{\Theta}^{T}(k) \zeta(\tau) - y(\tau))\nn \\
\ts \ts +\frac{1}{2}\tr[(\hat{\Theta}(k) - \Theta_{0})^{T} P_{0}^{-1} (\hat{\Theta}(k) -
\Theta_{0})],\;P_{0} = P_{0}^{T} \left(\in R^{60 \times 60}\right) >
0,\;\rho > 0.
\label{Jrobot}
\eea
This cost function $J_3$ is a modified version of the compact form of
$J_2$ in (\ref{J21}) with the estimates $\hat{A}_{21}$, $\hat{A}_{22}$
and $\hat{B}_2$ (of the parameters $A_{21}$, $A_{22}$ and $B_2$): 
\beq
J_2 = \sum_{\tau=k_0}^{k-1} (\hat{\Theta}^{T}(k) \zeta(\tau) -
y(\tau))^{T} (\hat{\Theta}^{T}(k) \zeta(\tau) - y(\tau)),
\label{J22}
\eeq
which is suitable for a non-iterative algorithm using a batch of
data \cite{o87}. The term $\frac{1}{2}\tr[(\hat{\Theta}(k) - \Theta_{0})^{T}
  P_{0}^{-1} (\hat{\Theta}(k) - \Theta_{0})]$ in (\ref{Jrobot}) represents a
penalty on the initial estimate $\hat{\Theta}_0$. The cost function
$J$ in (\ref{Jrobot}) is suitable for the derivation of an iterative
(adaptive) algorithm for the online estimate $\hat{\Theta}(k)$, using real-time
updated measurements (data) \cite{t03}.

The iterative algorithm for $\hat{\Theta}(k) = [\hat{\theta}_1(k),
  \hat{\theta}_2(k), \ldots, \hat{\theta}_{30}(k)] \in R^{60 \times
  30}$ as the estimate of $\Theta$ in the parametrized form 
(\ref{y(k)}) with measured signals $y(k)$ and $\zeta(k)$ has the form \cite{t03}:
\beq
\hat{\Theta}(k+1) = \hat{\Theta}(k) - \frac{P(k-1) \zeta(k)
  \epsilon^T(k)}{m^2(k)},\;k = k_0, k_0+1, k_0+2,\ldots,
\label{Thetak+1}
\eeq
with $\hat{\Theta}(k_0) = \Theta_0$ as a chosen initial estimate of
$\Theta$, where
\beq
\epsilon(k) = \hat{\Theta}^T(k) \zeta(k) - y(k)
\label{epsilon(k)}
\eeq
\beq
m(k) = \sqrt{\rho + \zeta^{T}(k)P(k-1)\zeta(k)},\;\rho > 0
\eeq
\beq
P(k) = P(k-1) - \frac{P(k-1) \zeta(k) \zeta^{T}(k)
P(k-1)}{m^{2}(k)},\;k = k_0, k_{0}+1, k_0+2, \ldots,
\label{Pk1}
\eeq
with $P(k_{0}-1) = P_{0} = P^{T}_{0} > 0$ chosen (typically, $P_0 >
p_0 I_{60}$ with a large value of $p_0 > 0$).

\mdp
This algorithm, with $\hat{\Theta}(k_0)$ and $P(k_{0}-1)$ chosen, generates
$\hat{\Theta}(k+1)$ (which depends on $P(k-1)$) and $P(k)$ (which is for the
next step $\hat{\Theta}(k+1)$, at each step $k = k_0, k_0+1, k_0+2,\ldots$,
iteratively.

\mdp
Some useful modifications \cite{t03} to this algorithm include: 

\begin{quote}
$\bullet$ gain matrix resetting: reset $P(k_{r}+1) = p_0 I$ whenever
 the minimum eigenvalue of $P(k_r)$, $\lambda_{\min}(P(k_r)) \leq p_1$
 for some small positive number $p_1 << p_0$, to avoid the gain matrix
 $P(k)$ to become close to being singular; and 

$\bullet$ parameter projection: force the components of
 $\hat{\Theta}(k)$ to stay in some intervals, to ensure robustness
 with respect to the approximation error $\eta_{(2)}(k)$ in (\ref{y(k)}).
\end{quote}

\medskip
Such a least-squares algorithm has some desired properties, see
\cite{t03} for details.

\subsubsection{Linear Model Identification Simulation}
\label{simulation}
To simulate the developed iterative linear model identification  
algorithm on the 5-robot system with nonlinear utility
functions, for determining the parameters $\hat{\Theta}^T(k) =
[\hat{A}_{21}(k), \hat{A}_{22}(k), \hat{B}_2(k)]$ online, at time $k$,
we first apply some chosen $U(k)$ to the $5$-robot system model
(\ref{Xequation3}): $X(k+1) = A X(k) + B U(k)$, 
to generate $X(k+1)$ as $z_{(1)}(k+1) = X(k+1)$ (see (\ref{z(1)})):
\beq
z_{(1)}(k+1) = [z_1(k+1), \ldots, z_{20}(k+1)]^T = X(k+1) \in R^{20},
\label{z(1)2}
\eeq
and use the utility functions (\ref{z(2)}) and (\ref{utility2}):
$z_{(2)}(k) = \psi_{(2)}(X(k))$, to calculate $z_{(2)}(k+1)$:
\beq
z_{(2)}(k+1) = \psi_{(2)}(X(k+1)).
\eeq
We then go to time $k+1$ for the next set of $z_{(1)}(\cdot) =
X(\cdot)$ and $z_{(2)}(\cdot)$. For this procedure, the initial
conditions are $z_{(1)}(k_0) = X(k_0)$ (and $z_{(2)}(k_0) =
\psi_{(2)}(X(k_0))$).

In this way, at time $k$, we have the measurements (data) of $y(k) =
z_{(2)}(k+1)$ and $\zeta(k)$ (from $z_{(1)}(k), z_{(2)}(k), U(k)$, see
(\ref{zeta})) in the linear model (\ref{y(k)}): 
$y(k) = \Theta^T \zeta(k) + \eta_{(2)}(k)$, 
for which the algorithm (\ref{Thetak+1})-(\ref{Pk1}) 
can be used for identifying/estimating the parameter matrix $\Theta$ containing
the parameter matrices $A_{21}$, $A_{22}$ and
$B_2$ of the linear model (\ref{z(2)(k+1)2}) with $\eta_{(2)}(k)$:
\beq
z_{(2)}(k+1) = A_{21} z_{(1)}(k) + A_{22} z_{(2)}(k) + B_2  U(k) +
\eta_{(2)}(k).
\label{z(2)(k+1)3}
\eeq

\mdp
To apply such an iterative algorithm for the linear model
estimation of the original nonlinear Koopman system model (\ref{xi(2)}), the
choice of $U(k)$ should be such that the robots make some
desired movements, rich enough to make the regressor signal $\zeta(k)$
in (\ref{zeta}) exciting enough to help the iterative algorithm to
estimate the parameter matrices $A_{21}, A_{22}, B_2$ of the above
model, and suitable enough to avoid collisions among the robots when
moving around their initial locations.

\subsubsection{Model Validation}
\label{validation}
For model validation, there are two sets of data to be generated and
collected: one
set from the original nonlinear model, and one set from the
approximate linear model.

\bigskip
{\bf Data from the original system model}. 
For the state vector $z(k) = [(z_{(1)}(k))^T,
  (z_{(2)}(k))^T]^T \in R^{50}$ in (\ref{z(k)}), the original $z_{(1)}$-subsystem
is (\ref{zisys}):
\beq
z_{(1)}(k+1) = A z_{(1)}(k) + B U(k),
\eeq
with $z_{(1)}(k) = X(k)$, plus the nonlinear model (\ref{xi(2)}):
\beq
z_{(2)}(k+1) = \xi_{(2)}(z_{(1)}(k), z_{(2)}(k), U(k)),
\label{nominal}
\eeq
where $z_{(2)}(k)$ is defined in (\ref{z(2)}):
\beq
z_{(2)}(k) = \psi_{(2)}(X(k)) = [\psi_{21}(X(k)), \ldots, 
\psi_{50}(X(k))]^T.
\eeq 
Although the function $\xi_{(2)}(\cdot, \cdot, \cdot)$ exists but it
is only nominal (and is hard to get), we can use 
\beq
z_{(2)}(k+1) = \psi_{(2)}(X(k+1)),
\label{z(2)(k+1)1}
\eeq 
to obtain $z_{(2)}(k+1)$. Hence, with the initial
conditions $z_{(1)}(k_0) = X(k_0)$, at time $k$, with the current $X(k)$ as
$z_{(1)}(k) = X(k)$ and a chosen $U(k)$, we generate $z_{(1)}(k+1) =
X(k+1)$, $z_{(2)}(k+1) = \psi_{(2)}(X(k+1))$, and then go to time
$k+1$, for the next set of $z_{(1)}(\cdot) = X(\cdot)$ and $z_{(2)}(\cdot)$. 

\bigskip
{\bf Data from the approximate linear model}. 
The approximate  (identified/estimated) linear system is of the form (\ref{zs1})
without $\eta_z(k)$:
\beq
\hat{z}(k+1) = \hat{A}_z \,\hat{z}(k) + \hat{B}_z\,U(k),
\eeq
(where $\hat{A}_z$ and $\hat{B}_z$ are the estimates of $A_z$ and
$B_z$) which is equal to the two equations:
\beq
z_{(1)}(k+1) = A z_{(1)}(k) + B U(k)
\label{z(1)(k+1)}
\eeq
\beq
\hat{z}_{(2)}(k+1) = \hat{A}_{21} z_{(1)}(k) + \hat{A}_{22} \hat{z}_{(2)}(k) + \hat{B}_2  U(k),
\label{z(2)(k+1)appr}
\eeq
where $\hat{A}_{21}$, $\hat{A}_{22}$ and
$\hat{B}_2$ are the final (fixed) estimates of $A_{21}$, $A_{22}$ and
$B_2$, obtained from the developed adaptive estimation algorithm,
after a system identification process is completed. 

This will generate a nominal linear approximation model for the
original nonlinear system. However, since such a linear model uses 
fixed parameter estimates $\hat{A}_{21}$, $\hat{A}_{22}$ and
$\hat{B}_2$, it may not capture the characteristics of the original
nonlinear system dynamically. 

\begin{rem}
\rm
An online (adaptive) linear model using adaptive parameters can be constructed as
\beq
z_{(1)}(k+1) = A z_{(1)}(k) + B U(k)
\label{z(1)(k+1)ol}
\eeq
\beq
\hat{z}_{(2)}(k+1) = \hat{A}_{21}(k) z_{(1)}(k) + \hat{A}_{22}(k) \hat{z}_{(2)}(k) + \hat{B}_2(k) U(k),
\label{z(2)(k+1)approl}
\eeq
where $\hat{A}_{21}(k)$, $\hat{A}_{22}(k)$ and
$\hat{B}_2(k)$ are the online (adaptive) estimates of $A_{21}$, $A_{22}$ and
$B_2$, obtained from the adaptive parameter estimation algorithm
(\ref{Thetak+1}) at each time instant $k$, in stead of the final
(fixed) estimates of $A_{21}$, $A_{22}$ and $B_2$, obtained after a system
identification process. 
\hspace*{\fill} $\Box$
\end{rem}

To use this approximate linear model to calculate $z_{(1)}(\cdot)$ and
$\hat{z}_{(2)}(\cdot)$, we set the initial conditions $z_{(1)}(k_0) = X(k_0)$
and $\hat{z}_{(2)}(k_0) = \psi_{(2)}(X(k_0))$, and, at time $k$, with the current
$z_{(1)}(k)$ and a chosen $U(k)$, we generate $z_{(1)}(k+1) = X(k+1)$
from (\ref{z(1)(k+1)}), and with the current $\hat{z}_{(2)}(k)$ and
$z_{(1)}(k)$, we generate $\hat{z}_{(2)}(k+1)$ from (\ref{z(2)(k+1)appr}),
and then go to time $k+1$, for the next set of $z_{(1)}(\cdot)$ and
$\hat{z}_{(2)}(\cdot)$. 

The signal $\hat{z}_{(2)}(\cdot)$ from (\ref{z(2)(k+1)appr}) is the
state vector of the approximate linear system (\ref{z(2)(k+1)appr}),
and is a functional approximation of $z_{(2)}(k) = \psi_{(2)}(X(k))$
from the utility functions (\ref{utility2}), as the state vector of the
nominal nonlinear model (\ref{nominal}).

\bp
{\bf Model validation}. 
To validate our proposed iterative identification algorithm for the
linear model (\ref{z(2)(k+1)3}), we calculate the values of
$z_{(1)}(\cdot) = X(\cdot)$, $z_{(2)}(\cdot)$ and $\hat{z}_{(2)}(\cdot)$, with some test
signals $U(k)$, from the above two schemes: 
$\hat{z}_{(2)}(k+1)$ from (\ref{z(2)(k+1)appr}) or
(\ref{z(2)(k+1)approl}),  and $z_{(2)}(k+1)$ from 
(\ref{z(2)(k+1)1}) (with $X(k+1)$ from (\ref{z(1)(k+1)})), and compare
their difference (the model validation error):
\beq
\tilde{z}_{(2)}(k+1) = \hat{z}_{(2)}(k+1) -
z_{(2)}(k+1),
\label{tildez(2)}
\eeq
to check the accuracy of the approximate linear model
(\ref{z(2)(k+1)appr}) for approximating the nonlinear model
(\ref{z(2)(k+1)1}) which has a nominal expression (\ref{nominal}).

\medskip
It is important that the test signals $U(k)$ for model validation
should be different from those used for the adaptive parameter
identification algorithm and should be diverse (with multiple sets of
$U(k)$ for multiple testings), to verify the suitability of the model
for broad system motions.

\bp
{\bf Error signals}. 
In this development, there are four notable error signals. 

\begin{quote}
$\bullet$ The first error
signal is the estimation error $\epsilon(k)$ in (\ref{epsilon(k)})
used in the iterative estimation algorithm (\ref{Thetak+1}), 
which is the difference between the known signal $\hat{A}_{21}(k)
z_{(1)}(k) + \hat{A}_{22}(k) z_{(2)}(k) + \hat{B}_2(k) U(k)$ and the
known signal $y(k) = z_{(2)}(k+1) = \psi_{(2)}(X(k+1))$:
\beq
\epsilon(k) = \hat{A}_{21}(k) z_{(1)}(k) +
\hat{A}_{22}(k) z_{(2)}(k) + \hat{B}_2(k) U(k) - z_{(2)}(k+1).
\label{epsilon(k)1}
\eeq
This error signal is available at each time instant $k$, 
because it is generated from known signals,
representing how close the estimated model output $\hat{y}(k) =
\hat{\Theta}^T(k) \zeta(k)$ to the real model output $y(k) =
z_{(2)}(k+1)$. Compared with (\ref{z(2)(k+1)appr}), 
$\hat{y}(k) = \hat{\Theta}^T(k) \zeta(k)$ is
\beq
\hat{y}(k) = \hat{A}_{21}(k) z_{(1)}(k) + \hat{A}_{22}(k) 
\hat{z}_{(2)}(k) + \hat{B}_2(k)  U(k),
\eeq
in which $\hat{A}_{21}(k)$, $\hat{A}_{22}(k)$ and $\hat{B}_2(k)$
explicitly depend on $k$, while in (\ref{z(2)(k+1)appr}), 
$\hat{A}_{21}$, $\hat{A}_{22}$ and $\hat{B}_2$ are obtained and 
fixed estimates of $A_{21}$, $A_{22}$ and $B_2$.

\medskip
$\bullet$ The second error signal is the modeling error $\eta_{(2)}(k)$ in
(\ref{z(2)(k+1)}) which is an unavailable signal in real time (because
$A_{21}$, $A_{22}$ and $B_2$ are nominal and unknown), representing how close 
 the nominal and unknown signal $A_{21} z_{(1)}(k) + A_{22} z_{(2)}(k) + B_2
 U(k)$ to the system signal $z_{(2)}(k+1) = \xi_{(2)}(z_{(1)}(k),
 z_{(2)}(k), U(k))
 = \psi_{(2)}(X(k+1))$ (see 
(\ref{nominal}) and (\ref{z(2)(k+1)1})). For one set of the estimates $(\hat{A}_{21},
 \hat{A}_{22}, \hat{B}_2)$ of $(A_{21}, A_{22}, B_2)$, there is one
 realized and known version $\hat{\eta}_{(2)}(k)$ of $\eta_{(2)}(k)$, based on
 the equation 
\beq
z_{(2)}(k+1) = \hat{A}_{21} z_{(1)}(k) + \hat{A}_{22} z_{(2)}(k) + \hat{B}_2  U(k) +
\hat{\eta}_{(2)}(k),
\label{hateta(2)(k)}
\eeq
that is, $- \hat{\eta}_{(2)}(k)$ is the difference between 
$\hat{A}_{21} z_{(1)}(k) + \hat{A}_{22} z_{(2)}(k) + \hat{B}_2 U(k)$ 
(for fixed or adaptive estimates $\hat{A}_{21}$, $\hat{A}_{22}$ and
$\hat{B}_2$) and $z_{(2)}(k+1) = \psi_{(2)}(X(k+1))$:
\beq
- \hat{\eta}_{(2)}(k) = \hat{A}_{21} z_{(1)}(k) + \hat{A}_{22}
z_{(2)}(k) + \hat{B}_2 U(k) - z_{(2)}(k+1).
\eeq
Hence, there are two versions of 
$\hat{\eta}_{(2)}(k)$, for fixed or adaptive estimates $\hat{A}_{21}$,
$\hat{A}_{22}$ and $\hat{B}_2$.

\medskip
One may compare the defined model (\ref{hateta(2)(k)}) with the
linear approximate model (\ref{z(2)(k+1)appr}), and 
the signal $- \hat{\eta}_{(2)}(k)$ in (\ref{hateta(2)(k)})
with the estimation error signal $\epsilon(k)$ in (\ref{epsilon(k)1}).

$\bullet$ The third error signal is the model validation error defined in (\ref{tildez(2)}):
\beq
\tilde{z}_{(2)}(k+1) = \hat{z}_{(2)}(k+1) -
z_{(2)}(k+1) = \hat{A}_{21} z_{(1)}(k) + \hat{A}_{22}
\hat{z}_{(2)}(k) + \hat{B}_2  U(k) - z_{(2)}(k+1),
\eeq
where $\hat{z}_{(2)}(k+1) = \hat{A}_{21} z_{(1)}(k) + \hat{A}_{22}
\hat{z}_{(2)}(k) + \hat{B}_2  U(k)$ is the linear system approximation
of the nonlinear signal $z_{(2)}(k+1) =
\xi_{(2)}(z_{(1)}(k), z_{(2)}(k), U(k)) = \psi_{(2)}(X(k+1))$, either
fixed parameter estimates $\hat{A}_{21}$, $\hat{A}_{22}$ and
$\hat{B}_2$ or adaptive (online) parameter estimates 
$\hat{A}_{21} = \hat{A}_{21}(k)$, $\hat{A}_{22} = \hat{A}_{22}(k)$ and
$\hat{B}_{2} = \hat{B}_2(k)$. 

Hence, there are two versions of the  model validation error
$\tilde{z}_{(2)}(k+1)$.

$\bullet$ The fourth error signal is the {\it a posteriori} estimation
error
\beq
\epsilon_a(k) = \hat{\Theta}^T(k+1) \zeta(k) - y(k),
\label{epsilona(k)}
\eeq
as compared with the estimation error $\epsilon(k) = \hat{\Theta}^T(k)
\zeta(k) - y(k)$ defined in (\ref{epsilon(k)}). The error
$\epsilon_a(k)$ is a reconstructed version of $\epsilon(k)$, using the
updated parameter estimate $\hat{\Theta}(k+1)$, and 
$\epsilon_a(k)$ is a meaningful measure of the parameter estimation
performance. If $\hat{\Theta}(k)$ converges to a constant matrix (or 
$\lim_{k \rightarrow \infty} (\hat{\Theta}(k+1) - \hat{\Theta}(k)) =
0$, a weaker condition), then $\epsilon_a(k)$ converges to
$\epsilon(k)$. Generically, $\epsilon_a(k)$ is smaller than
$\epsilon(k)$ in magnitude, due to the estimation improvement by the
updated parameter estimate $\hat{\Theta}(k+1)$.
\end{quote}

\medskip
Based on (\ref{hateta(2)(k)}) with adaptive (online) parameter estimates 
 $\hat{A}_{21}(k)$, $\hat{A}_{22}(k)$ and $\hat{B}_2(k)$, we can have
the adaptive (online) linear Koopman system model:
\beq
z_{(2)}(k+1) = \hat{A}_{21}(k) z_{(1)}(k) + \hat{A}_{22}(k) z_{(2)}(k) + \hat{B}_2(k) U(k)
\label{hateta(2)(k)1}
\eeq

The model validation and error signal analysis are also applicable to
the bilinear model case.

\subsection{Bilinear Model Identification}
We first recall the Taylor series expansion (\ref{fi}) of the
component $f_i(x)$ of a generic vector function $f(x) = [f_1(x),
  f_2(x), \ldots, f_m(x)]^T \in R^m$ of $x = [x_1, x_2, \ldots, x_n]^T
\in R^n$.

\bigskip
{\bf Model setup}. 
To be relevant to our robot system, we let 
\beq
x = [z_{(1)}^T, z_{(2)}^T, U^T]^T,
\eeq
where $z_{(1)} = z_{(1)}(k)$, $z_{(2)} = z_{(2)}(k)$ and $U = U(k)$
are the robot system variables, Koopman system variables and control
variables, respectively. Corresponding to 
$x = [z_{(1)}^T, z_{(2)}^T, U^T]^T$, we choose
\beq
x_0 = [z_{(1)0}^T, z_{(2)0}^T, U_0^T]^T,\;U_0 = 0_{1 \times 6}
\footnote{For a bilinear model formulation, we consider the system with 3
  robots, to keep the system vector (and the corresponding system
  matrix) dimensions moderate: $z_{(1)}(k) \in R^{12}$, $z_{(2)}(k) \in R^{18}$ and $U(k) \in R^6$.},
\eeq
where $z_{(1)0}$ is the desired state vector of the robot system, and 
$z_{(2)0}$ is the desired state vector of the Koopman system when the
desired $z_{(1)0}$ is achieved by $z_{(1)}$. 

With $x = [z_{(1)}^T, z_{(2)}^T, U^T]^T$ and, accordingly, 
\beq
H_i = H_{f_i}(x_0) = \left[ \begin{array}{ccc}
H_{i,11} & H_{i,12} & H_{i,13} \\
H_{i,21} & H_{i,22} & H_{i,23} \\
H_{i,31} & H_{i,32} & H_{i,33},
\end{array}
\right] = H_i^T,
\label{Hi}
\eeq
since $U - U_0 = U$ is expected to converge to $0_{1 \times 6}$, we
may ignore the term $U^T H_{i,33} U$ in $(x - x_0)^T H_{f_i}(x_0) (x -
x_0)$. Thus, we can approximate $f_i$ in (\ref{fi}) as
\bea
f_i(x) \ts \approx \ts f_i(x_{0}) + (\nabla f_i(x_0))^T (x - x_0) + \frac{1}{2} (x -
x_0)^T H_{f_i}(x_0) (x - x_0)\; \mbox{without $U^T H_{i,33} U$}.
\label{fi1}
\eea

Excluding the term $U^T H_{i,33} U$ from $f_i(x)$ will lead to a
bilinear model linear (more precisely, affine) in the control $U(k)$,
which is useful for reaching the goal of simplifying the solution
procedure for the optimal control problem: from maximizing a set of
nonlinear functions to maximizing a set of linear or affine functions,
the function sets both containing the control signal $U(k)$.

\begin{rem}
\rm
The linear model studied so far is based on the
approximation
\beq
f_i(x) \approx (\nabla f_i(x_0))^T (x - x_0),\;x_0 = 0_{1 \times n},
\label{lineara}
\eeq
which, when applied to the robot system with nonlinear utility
functions, may not provide enough capacity to solve the control
problem of driving the robots to their desired positions.

In this new formulation, the approximation scheme (\ref{fi1}) is to be
used to derive a
parametrized model which contains the off-set term $f_i(x_{0})$, the
perturbed term $x - x_0$ with nonzero $x_0$, bilinear terms of 
$[(z_{(1)} - z_{(1)0})^T, (z_{(2)} - z_{(2)0})^T]^T$ 
and $U$, and quadratic terms of $[(z_{(1)} - z_{(1)0})^T, (z_{(2)} -
  z_{(2)0})^T]^T$, in addition to linear terms from $(\nabla
f_i(x_0))^T (x - x_0)$, with $x = [z_{(1)}^T, z_{(2)}^T, U^T]^T$. Such
a new bilinear scheme is expected to provide more capacity in generating more
accurate approximation of the nonlinear utility functions and more
powerful control signals to maximize the utility functions, to drive
the robots to their desired locations. 
\hspace*{\fill} $\Box$
\end{rem}

\medskip
{\bf Model parametrization}. 
While the terms $f_i(x_{0})$, $\nabla f_i(x_0)$ and $H_{f_i}(x_0)$
(\ref{fi1}) may be calculated by definition when the function $f_i(x)$
is known, the error 
\beq
e_i = f_i(x) - f_i(x_{0}) + (\nabla f_i(x_0))^T (x - x_0) + \frac{1}{2} (x -
x_0)^T H_{f_i}(x_0) (x - x_0)\; \mbox{without $U^T H_{i,33} U$}
\eeq
may not be always minimized by the calculated 
$f_i(x_{0})$, $\nabla f_i(x_0)$ and $H_{f_i}(x_0)$. 

\medskip
A minimization-based method can be used to calculate 
$f_i(x_{0})$, $\nabla f_i(x_0)$ and $H_{f_i}(x_0)$ as parameters,
based on some cost function similar to that given (\ref{J22}). 

\medskip
An iterative (adaptive) algorithm is desirable to estimate 
$f_i(x_{0})$, $\nabla f_i(x_0)$ and $H_{f_i}(x_0)$ (and some
accumulative effect of those higher-order terms in (\ref{fi})), with
updated measurements (data). Such an online algorithm can be derived,
by minimizing some cost function similar to that given in
(\ref{Jrobot}), of the form (\ref{Thetak+1}), for which some model
parametrization preparation is needed.

\medskip
Based on the new approximation scheme (\ref{fi1}), for the 
Koopman system model
\beq
z_{(2)}(k+1) = \xi_{(2)}(z_{(1)}(k), z_{(2)}(k), U(k))
 \stackrel{\triangle} = f(x) = [f_1(x), \ldots, f_i(x), \ldots]^T,
\label{f(x)}
\eeq
with the notation $z_{(2)}(k) =
[z_{(2),1}(k), z_{(2),2}(k), \ldots, z_{(2),i}(k), \ldots, ]^T$, 
a bilinear model based on
\beq
z_{(2),i}(k+1) \approx f_i(x_{0}) + (\nabla f_i(x_0))^T (x - x_0) + \frac{1}{2} (x -
x_0)^T H_{f_i}(x_0) (x - x_0)\; \mbox{without $U^T H_{i,33} U$},
\label{bilinear}
\eeq
is to be derived and estimated for its parameter matrices $f_i$, 
$\nabla f_i$ and $H_{f_i}$ without $H_{i,33}$, for which
a parametrized model (in terms of the parameters of $f_i(x_{0})$,
$\nabla f_i(x_0)$ and
$H_{f_i}(x_0)$) and an adaptive algorithm (to estimate these
parameters) will be developed.

\medskip
For $z_{(2)}(k) = [z_{(2),1}(k), z_{(2),2}(k), \ldots, z_{(2),i}(k),
  \ldots]^T$, applying a parametrization procedure (of separating
unknown parameters from known signals) on (\ref{bilinear}), we can
derive the bilinear model
\bea
z_{(2)}(k+1) \ts = \ts A_{21} (z_{(1)}(k) - z_{(1)0}) + A_{22}
(z_{(2)}(k)-z_{(2)0}) + B_2 U(k) \nn\\
\ts \ts + f(z_{(1)0}, z_{(2)0}) + \Phi_{1}^T g_{1}(z_{(1)}(k)-z_{(1)0}) \nn\\
\ts \ts + \Phi_{2}^T g_{2}(z_{(2)}(k)-z_{(2)0}, z_{(1)}(k)-z_{(1)0})
+ \Phi_{3}^T g_{3}(z_{(2)}(k)-z_{(2)0}) \nn \\
\ts \ts  + \Phi_{4}^T g_{4}
(z_{(1)}(k)-z_{(1)0}, U(k)) + \Phi_{5}^T g_{5}
(z_{(2)}(k)-z_{(2)0}, U(k)) + \eta_{(2)}(k),
\label{pm}
\eea
for some unknown parameter matrices $A_{21}$, $A_{22}$, $B$, $\Phi_i$, 
$i=1,2,\ldots, 5$, and vector $f(z_{(1)0}, z_{(2)0})$ to be estimated,
and some known (available) vector signals $g_i$, $i=1,2,\ldots, 5$,
where $\eta_{(2)}(k)$ represents the modeling error. 
In this model, $A_{21} (z_{(1)}(k) - z_{(1)0}) + A_{22}
(z_{(2)}(k)-z_{(2)0}) + B_2 U(k)$ is the linear part at the operating
point $(z_{(1)0}, z_{(2)0})$,  
$f(z_{(1)0}, z_{(2)0})$ is the operating point dynamics offset term,
and other terms are from the bilinear modeling, in particular, their
$i$th rows are
\beq
(\Phi_{1}^T g_{1}(z_{(1)}(k)-z_{(1)0}))_i = 0.5 
(z_{(1)}(k)-z_{(1)0})^T H_{i,11} (z_{(1)}(k)-z_{(1)0})
\eeq
\beq
(\Phi_{2}^T g_{2}(z_{(2)}(k)-z_{(2)0}, z_{(1)}(k)-z_{(1)0}))_i = 
(z_{(2)}(k)-z_{(2)0})^T H_{i,21} (z_{(1)}(k)-z_{(1)0})
\eeq
\beq
(\Phi_{3}^T g_{3}(z_{(2)}(k)-z_{(2)0}))_i = 
0.5 (z_{(2)}(k)-z_{(2)0})^T H_{i,22} (z_{(2)}(k)-z_{(2)0})
\eeq
\beq
(\Phi_{4}^T g_{4} (z_{(1)}(k)-z_{(1)0}, U(k)))_i = 
(z_{(1)}(k)-z_{(1)0})^T H_{i,13}U(k)
\eeq
\beq
(\Phi_{5}^T g_{5} (z_{(2)}(k)-z_{(2)0}, U(k)))_i = 
(z_{(2)}(k)-z_{(2)0})^T H_{i,23}U(k),
\eeq
for $H_{i,jk}$ being the parts of $H_i$ in (\ref{Hi}). To
further clarify the parametrized model (\ref{pm}), we note that, since
$H_i = H_i^T$, that is, $H_{i,21}^T = H_{i,12}$, $H_{i,31}^T =
H_{i,13}$ and $H_{i,32}^T = H_{i,23}$, we have
\beq
U^T(k) H_{i,31}(z_{(1)}(k)-z_{(1)0}) = 
(z_{(1)}(k)-z_{(1)0})^T H_{i,31}^T U(k) = 
(z_{(1)}(k)-z_{(1)0})^T H_{i,13}U(k) 
\eeq
\beq
U^T(k) H_{i,32} (z_{(2)}(k)-z_{(2)0}) = 
(z_{(2)}(k)-z_{(2)0})^T H_{i,32}^T U(k) = 
(z_{(2)}(k)-z_{(2)0})^T H_{i,23}U(k) 
\eeq
\beq
(z_{(2)}(k)-z_{(2)0})^T H_{i,21} (z_{(1)}(k)-z_{(1)0}) = 
(z_{(1)}(k)-z_{(1)0})^T H_{i,12} (z_{(2)}(k)-z_{(2)0}).
\eeq

Also, from (\ref{f(x)}) and the definition of $z_{(2)}(k) =
\psi_{(2)}(X(k))$ whose components are the nonlinear utility functions
(see (\ref{z(2)})), we have
\beq
f(x_0) = f(z_{(1)0}, z_{(2)0}) = \psi_{(2)}(A X(0) + B U(0)) = \psi_{(2)}(A z_{(1)0} + B U_0)
= \psi_{(2)}(A z_{(1)0}).
\eeq

As a comparison to (\ref{pm}), the bilinear model
proposed in \cite{bfv21} has the form:
\bea
z_{(2)}(k+1) \ts = \ts A_{21} z_{(1)}(k) + A_{22}
z_{(2)}(k) + B_2 U(k) \nn\\
\ts \ts  + \Phi_{4}^T g_{4}
(z_{(1)}(0), U(k)) + \Phi_{5}^T g_{5}
(z_{(2)}(0), U(k)).
\eea

\medskip
{\bf Model adaptation}. 
The model (\ref{pm}) can be written in the form (\ref{y(k)}): 
\beq
y(k) = \Theta^T \zeta(k) + \eta_{(2)}(k),
\label{y(k)1}  
\eeq
for an unknown parameter matrix $\Theta$ to be estimated and a known vector
signal $\zeta(k)$:
\bea
\label{Thetab}
\Theta \ts = \ts [A_{21}, A_{22}, B_2, f(z_{(1)0}, z_{(2)0}),
  \Phi_{1}^T, \Phi_{2}^T, \Phi_{3}^T, \Phi_{4}^T, \Phi_{5}^T]^T \\
\zeta(k)\ts = \ts [(z_{(1)}(k) - z_{(1)0})^T, (z_{(2)}(k)-z_{(2)0})^T,
  (U(k))^T, 1, (g_{1}(z_{(1)}(k)-z_{(1)0}))^T, \nn\\
\ts \ts 
(g_{2}(z_{(2)}(k)-z_{(2)0}, z_{(1)}(k)-z_{(1)0}))^T,
  (g_{3}(z_{(2)}(k)-z_{(2)0}))^T, (g_{4}(z_{(1)}(k)-z_{(1)0},
  U(k)))^T,\nn\\
\ts \ts (g_{5}(z_{(2)}(k)-z_{(2)0}, U(k)))^T]^T,
\eea
so that an iterative algorithm for estimating
$\Theta$ can be developed, similar to that for the linear model case
in Section \ref{iterative}. More specifically, with 
$\hat{A}_{21}$, $\hat{A}_{22}$, $\hat{B}_2$, $\hat{f}_0$, $\hat{\Phi}_{1}^T$, $\hat{\Phi}_{2}^T$,
$\hat{\Phi}_{3}^T$, $\hat{\Phi}_{4}^T$ and $\hat{\Phi}_{5}^T$ being
the estimates of the unknown parameters 
$A_{21}$, $A_{22}$, $B_2$, $f_0 = f(z_{(1)0}, z_{(2)0})$,
$\Phi_{1}^T$, $\Phi_{2}^T$, $\Phi_{3}^T$, $\Phi_{4}^T$ and
$\Phi_{5}^T$, respectively, we can define the estimation error as
\bea
\epsilon(k) \ts = \ts \hat{A}_{21}(k) (z_{(1)}(k) - z_{(1)0}) + \hat{A}_{22}(k)
(z_{(2)}(k)-z_{(2)0}) + \hat{B}_2(k) U(k) \nn\\
\ts \ts + \hat{f}_0(k) + \hat{\Phi}_{1}^T(k) g_{1}(z_{(1)}(k)-z_{(1)0}) \nn\\
\ts \ts + \hat{\Phi}_{2}^T(k) g_{2}(z_{(2)}(k)-z_{(2)0}, z_{(1)}(k)-z_{(1)0})
+ \hat{\Phi}_{3}^T(k) g_{3}(z_{(2)}(k)-z_{(2)0}) \nn \\
\ts \ts  + \hat{\Phi}_{4}^T(k) g_{4}
(z_{(1)}(k)-z_{(1)0}, U(k)) + \hat{\Phi}_{5}^T(k) g_{5}
(z_{(2)}(k)-z_{(2)0}, U(k)) - z_{(2)}(k+1),
\eea
which can be written in the form (\ref{epsilon(k)}): 
$\epsilon(k) = \hat{\Theta}^T(k) \zeta(k) - y(k)$. Then, the iterative
(adaptive) algorithm (\ref{Thetak+1})--(\ref{Pk1}) can be used to
generate the estimate $\hat{\Theta}(k)$ of $\Theta$.

\bp
{\bf Model simulation and validation}. The simulation procedure in
Section \ref{simulation} and validation procedure in Section
\ref{validation} for the linear model case can be similarly followed
to conduct simulation and validation of the bilinear model developed
in this subsection. The three error signals discussed in Section
\ref{validation} also have their similar meanings for a bilinear model.

\bp
{\bf Online Koopman model}. The adaptive (online) bilinear Koopman model:
\bea
z_{(2)}(k+1) \ts = \ts \hat{A}_{21}(k) (z_{(1)}(k) - z_{(1)0}) + \hat{A}_{22}(k)
(z_{(2)}(k)-z_{(2)0}) + \hat{B}_2(k) U(k) \nn\\
\ts \ts + \hat{f}_0(k) + \hat{\Phi}_{1}^T(k) g_{1}(z_{(1)}(k)-z_{(1)0}) \nn\\
\ts \ts + \hat{\Phi}_{2}^T(k) g_{2}(z_{(2)}(k)-z_{(2)0}, z_{(1)}(k)-z_{(1)0})
+ \hat{\Phi}_{3}^T(k) g_{3}(z_{(2)}(k)-z_{(2)0}) \nn \\
\ts \ts  + \hat{\Phi}_{4}^T(k) g_{4}
(z_{(1)}(k)-z_{(1)0}, U(k)) + \hat{\Phi}_{5}^T(k) g_{5}
(z_{(2)}(k)-z_{(2)0}, U(k)) 
\label{z2k+1b}
\eea
will be used to approximate the nonlinear utility functions
(\ref{utility2}) as (\ref{lf}), as the adaptive (online) version of 
(\ref{pm}) without $\eta_{(2)}(k)$, to be used for solving the control
problem to be formulated.

\subsection{General Framework of Koopman System Approximation}
For a generic vector function $f(x) = [f_1(x), f_2(x), \ldots,
  f_m(x)]^T \in R^m$ of $x = [x_1, x_2, \ldots, x_n]^T \in R^n$, the
second-order Taylor series expansion of $f_i(x)$ in (\ref{fi}) at $x_0
= [x_{01}, \ldots, x_{0n}]^T$ can be written as
\beq
f_i(x) = f_i(x_{0}) + \sum_{j=1}^n \frac{\partial f_i(x_0)}{\partial
  x_j} (x_j - x_{0j}) + \frac{1}{2!} \sum_{j=1}^n \sum_{p=1}^n 
\frac{\partial^2 f_i(x_0)}{\partial
  x_j \partial x_p} (x_j - x_{0j})(x_p - x_{0p}) + \mbox{h.o.t.}.
\label{fi2}
\eeq

\medskip
{\bf Higher-order approximation of Koopman systems}. We may also look
into the third-order Taylor series expansion of $f_i(x)$:
\bea
f_i(x) \ts = \ts f_i(x_{0}) + \sum_{j=1}^n \frac{\partial f_i(x_0)}{\partial
  x_j} (x_j - x_{0j}) 
+ \frac{1}{2!} \sum_{j=1}^n \sum_{p=1}^n \frac{\partial^2 f_i(x_0)}{\partial
  x_j \partial x_p} (x_j - x_{0j})(x_p - x_{0p}) \nn\\
\ts \ts  + 
\frac{1}{3!} \sum_{j=1}^n \sum_{p=1}^n \sum_{l=1}^n 
\frac{\partial^3 f_i(x_0)}{\partial
  x_j \partial x_p \partial x_l} (x_j - x_{0j})(x_p - x_{0p})(x_l 
- x_{0l}) + \mbox{h.o.t.}, 
\label{fi3}
\eea
or even higher-order Taylor series expansion of $f_i(x) =
z_{(2),i}(k+1) = \psi_i(X(k+1)) = \xi_{(2),i}(z(k), u(k))$, to build
some higher-order (multilinear) approximation models of a Koopman
system. Such a higher-order approximation model can also be expressed
in a parametrized form (\ref{y(k)}) or (\ref{y(k)1}) based on which an
adaptive algorithm can be implemented to estimate its parameters. 

Higher-order approximations can capture more essence of the nonlinear
functions $f_i(x(k)) = \xi_{(2),i}(z(k), u(k))$, which is desirable
and crucial for many applications including multi-robot control.

\medskip
Recall that the robot system control problem of this paper is to
maximize some nonlinear utility functions as the Koopman system
variables $z_{(2),i}(k)$, by transforming a nonlinear programming
problem to a linear programming problem. Hence, for such problems, the
higher-order approximation Koopman system models (the 
$z_{(2)}(k)$-subsystem) should not contain higher-order terms of the
control inputs $U(k)$ (but the linear or affine terms of $U(k)$) in the
approximation expression of $z_{(2),i}(k+1)$, in order to facilitate
the control design to find $U(k)$ to maximize the utility functions
$u_i(k+1)$ involving $z_{(2)}(k+1)$, using a linear programming
framework for the nonlinear programming framework of
maximizing the original nonlinear utility functions.

\bp
{\bf Online Koopman system approximation procedure}. For the
multi-robot system (\ref{z(1)(k+1)0}) with nonlinear utility functions
(\ref{utility2}) whose components $\phi_{i}^{(j)}(X(k))$ are modeled
as the state variables $z_{(2)}(k)$ of the (nominal) nonlinear Koopman
system (\ref{xi(2)}):
\beq
z_{(2)}(k+1) = \xi_{(2)}(z_{(1)}(k), z_{(2)}(k), U(k)),
\label{xi(2)2}
\eeq
a linear model (\ref{z(2)(k+1)}) or a bilinear model (\ref{pm}) can be
used for this nonlinear system, both leading to a
parametrized model of the form (\ref{y(k)}) or (\ref{y(k)1}):
\beq
y(k) = \Theta^T \zeta(k) + \eta_{(2)}(k).
\eeq  

Based on this parametrized model, an estimation error 
$\epsilon(k) = \hat{\Theta}^T(k) \zeta(k) - y(k)$ is formed, where 
$\hat{\Theta}(k)$ is the estimate of the unknown $\Theta$, which is
adaptively updated from an iterative (online) algorithm of the form 
(\ref{Thetak+1})--(\ref{Pk1}). Then, an adaptive (online) linear or
bilinear approximation model of (\ref{xi(2)2}) can be obtained as 
(\ref{hateta(2)(k)1}) or (\ref{z2k+1b}) which can be used for control
design.

\setcounter{equation}{0}
\section{Optimal Control Formulation with Koopman Models}
The optimal control problem is stated as: for the robot system
(\ref{Xequation3}):
\beq
X(k+1) = A X(k) + B U(k),
\eeq
find the control input $U(k)$ which maximizes the total utility
function
\beq
u(k+1) = \sum_{i=1}^6 u_i(k+1)
\eeq
for $u_i(k)$ given in (\ref{utility2}) or (\ref{lf}):
\beq
u_{i}(k) = \sum_{j=1}^6 \omega_i^{(j)} \phi_{i}^{(j)}(X(k)) = 
\sum_{j=1}^6 \omega_i^{(j)} z_{20+6(i-1)+j}(k),\;i = 1,2,\cdots,5,
\eeq
subject to the constraints: $U_i(k) \in (-4, 3)$, for 
$U(k) = [U_1(k), \ldots, U_{10}(k)]^T$: $U(k) \in
U_c$ for 
\beq
U_c = \{U(k)\;|\;U_i(k) \in (-4, 3), i=1,\ldots, 10\}.
\eeq

With $z_{(2)}(k) = [z_{21}(k), \ldots, z_{50}(k)]^T = 
[(z_{(2)1}(k))^T, \ldots, (z_{(2)5}(k))^T]^T \in
R^{30}$ for $z_{(2)i}(k) \in R^6$, $\omega_i = [\omega_i^{(1)}, \ldots,
  \omega_i^{(6)}]^T \in R^6$, $i = 1,\cdots,5$, and $\omega =
[\omega_1^T, \ldots, \omega_5^T]^T \in R^{30}$, 
we can write $u(k+1)$ as
\beq
u(k+1) = \sum_{i=1}^6 \omega_i^{T} z_{(2)i}(k+1) = \omega^T z_{(2)}(k+1),
\eeq
ans denote the optimal control solution as
\beq
U^*(k) = {\rm arg \; max}_{\;U(k) \in U_c\;} \;
\omega^T z_{(2)}(k+1).
\label{U*kb}
\eeq

In the original optimal control problem, we need to solve the control
problem with
\beq
z_{(2)}(k+1) =
\psi_{(2)}(X(k+1)) = \psi_{(2)}(A X(k) + B U(k)) (=
\xi_{(2)}(z_{(1)}(k), z_{(2)}(k), U(k))),
\eeq
for the given nonlinear functions $\psi_{(2)}(\cdot)$ (see 
(\ref{utility2}), (\ref{zij}), (\ref{zig2}), (\ref{lf}) and
(\ref{z(2)})) which are nonlinear in nature (and so are the nominal
(virtual) functions $\xi_{(2)}(z_{(1)}(k), z_{(2)}(k), U(k))$).

Using the developed approximation-based Koopman system modeling
techniques, we can transform the this nonlinear programming problem to
the linear programming problem of solving (\ref{U*kb}), with
$z_{(2)}(k+1)$ from a linear or bilinear model developed.

\bigskip
{\bf Solution with a linear Koopman model}. In this case, the adaptive
linear model is
\beq
z_{(2)}(k+1) = \hat{A}_{21}(k) z_{(1)}(k) + \hat{A}_{22}(k) z_{(2)}(k)
+ \hat{B}_2(k) U(k),
\eeq
and we find $U(k)$ to maximize $\omega^T z_{(2)}(k+1)$, for $U(k) \in
U_c$, that is,
\begin{equation}
U^*(k) = \arg \max _{U(k)\in U_c} w^T(\hat{A}_{21}(k)z_{(1)}(k) +
\hat{A}_{22}(k)z_{(2)}(k)+\hat{B}_2(k)U(k)).
\eeq
Thus, this is a linear programming problem:
\begin{equation}
\begin{array}{ll@{}ll}
{\rm Maximize}  & 
w^T(\hat{A}_{21}(k)z_{(1)}(k)+\hat{A}_{22}(k)z_{(2)}(k)+\hat{B}_2(k)U(k)),&\\*[0.05in]
& {\rm subject\; to}
-4\leq U_i(k)\leq 3, \; i\in\{1,2,3,4,5,6\},
\end{array}
\label{LP_base}
\end{equation}

In Matlab, there is a built function $\mathtt{linprog}$ can solve such
linear programming problems. 

\bigskip
{\bf Solution with a bilinear Koopman model}. In this case, the
adaptive bilinear model for $z_{(2)}(k+1)$ is given in (\ref{z2k+1b})
in which the $i$th row of 
$\hat{\Phi}_{4}^T(k) g_{4} (z_{(1)}(k)-z_{(1)0}, U(k)) \in R^{18}$ and 
$\hat{\Phi}_{5}^T(k) g_{5} (z_{(2)}(k)-z_{(2)0}, U(k)) \in R^{18}$ should be
expressed explicitly in terms of $U(k)$:
\beq
(\hat{\Phi}_{4}^T(k) g_{4} (z_{(1)}(k)-z_{(1)0}, U(k)))_i = 
(z_{(1)}(k)-z_{(1)0})^T \hat{H}_{i,13}(k) U(k) 
\eeq
\beq
(\hat{\Phi}_{5}^T(k) g_{5} (z_{(2)}(k)-z_{(2)0}, U(k)))_i = 
(z_{(2)}(k)-z_{(2)0})^T \hat{H}_{i,23}U(k), 
\eeq
in order to explicitly find $U(k)$ to solve the maximization
problem. Hence, with 
\beq
\hat{G}_4(k) = 
\left[ \begin{array}{c}
(z_{(1)}(k)-z_{(1)0})^T \hat{H}_{1,13}(k) \\
(z_{(1)}(k)-z_{(1)0})^T \hat{H}_{2,13}(k) \\
\vdots\\
(z_{(1)}(k)-z_{(1)0})^T \hat{H}_{18,13}(k) 
\end{array}
\right]
\eeq
\beq
\hat{G}_5(k) = 
\left[ \begin{array}{c}
(z_{(2)}(k)-z_{(2)0})^T \hat{H}_{1,23}(k)\\
(z_{(2)}(k)-z_{(2)0})^T \hat{H}_{2,23}(k)\\
\vdots\\
(z_{(2)}(k)-z_{(2)0})^T \hat{H}_{18,23}(k)
\end{array}
\right]
\eeq
for the 3-robot case, we have the adaptive model
\bea
z_{(2)}(k+1) \ts = \ts \hat{A}_{21}(k) (z_{(1)}(k) - z_{(1)0}) + \hat{A}_{22}(k)
(z_{(2)}(k)-z_{(2)0}) + \hat{f}_0(k) + \hat{\Phi}_{1}^T(k)
g_{1}(z_{(1)}(k)-z_{(1)0}) \nn\\
\ts \ts + \hat{\Phi}_{2}^T(k) g_{2}(z_{(2)}(k)-z_{(2)0}, z_{(1)}(k)-z_{(1)0})
+ \hat{\Phi}_{3}^T(k) g_{3}(z_{(2)}(k)-z_{(2)0}) \nn \\
\ts \ts  + (\hat{B}_2(k) + \hat{G}_4(k) + \hat{G}_5(k))U(k),
\eea
for solving the linear programming version of the problem (\ref{U*kb}):
\beq
U^*(k) = {\rm arg \; max}_{\;U(k) \in U_c\;} \;
\omega^T z_{(2)}(k+1).
\eeq

\medskip
Hence, with either a linear model or a bilinear model of the utility
functions $z_{(2)}(k+1)$, the optimal control solution procedure
becomes much simpler than that with maximizing the original nonlinear
utility functions, as both models are either linear or affine in the
control signal $U(k)$.

\section{Concluding Remarks}
In this paper we have studied some key techniques for Koopman system
approximation based optimal control of multiple robots with nonlinear
utility functions which are to be maximized to generate the control
signal for desired system tracking performance. Essentially, the
Koopman operator, when applied on a set of observation functions
of the state vector of a nonlinear dynamic system $x(k+1) = f(x(k),
u(k))$, generates a set of dynamic equations which, through a
transformation, lead to a set of dynamic system equations which can be
approximated by linear, bilinear or multilinear equations to
facilitate designs for application problems.

In particular, to develop such a framework,
 we established the functional transformations (\ref{nnon3})
  which transform the Koopman equations (\ref{Nobsc}) to a Koopman
  dynamic system (\ref{zsysc1}); we developed an iterative (adaptive)
  algorithm for updating the linear Koopman system model
  (\ref{nsys2}), using incoming signal measurements (data), to build
  an online linear Koopman system model; we proposed a bilinear model
  approximation scheme (\ref{fi1}) and applied it to the nonlinear
  Koopman system model (\ref{f(x)}) to derive a parametrized bilinear
  Koopman system model (\ref{pm}), and formed an iterative (adaptive)
  bilinear Koopman system (\ref{z2k+1b}), to be updated using incoming
  data; and we also transferred the original nonlinear
  programming control problem to a linear programming control problem,
  using the linear or bilinear Koopman system model.

\bp
{\bf Advantage of a bilinear model over a linear model}. The bilinear
model (\ref{z2k+1b}):
\bea
z_{(2)}(k+1) \ts = \ts \hat{A}_{21}(k) (z_{(1)}(k) - z_{(1)0}) + \hat{A}_{22}(k)
(z_{(2)}(k)-z_{(2)0}) + \hat{B}_2(k) U(k) \nn\\
\ts \ts + \hat{f}_0(k) + \hat{\Phi}_{1}^T(k) g_{1}(z_{(1)}(k)-z_{(1)0}) \nn\\
\ts \ts + \hat{\Phi}_{2}^T(k) g_{2}(z_{(2)}(k)-z_{(2)0}, z_{(1)}(k)-z_{(1)0})
+ \hat{\Phi}_{3}^T(k) g_{3}(z_{(2)}(k)-z_{(2)0}) \nn \\
\ts \ts  + \hat{\Phi}_{4}^T(k) g_{4}
(z_{(1)}(k)-z_{(1)0}, U(k)) + \hat{\Phi}_{5}^T(k) g_{5}
(z_{(2)}(k)-z_{(2)0}, U(k)) 
\eea
is based on the bilinear approximation (\ref{fi1}): 
\bea
f_i(x) \ts \approx \ts f_i(x_{0}) + (\nabla f_i(x_0))^T (x - x_0) + \frac{1}{2} (x -
x_0)^T H_{f_i}(x_0) (x - x_0)\; \mbox{without $U^T H_{i,33} U$},
\eea
which has some additional key components than the linear model
(\ref{hateta(2)(k)1}):
\beq
z_{(2)}(k+1) = \hat{A}_{21} z_{(1)}(k) + \hat{A}_{22} z_{(2)}(k) + \hat{B}_2  U(k)
\eeq
which is based on the linear approximation (\ref{lineara}):
\bea
f_i(x) \ts \approx \ts (\nabla f_i(x_0))^T (x - x_0).
\eea
Thus, a bilinear model has more capacity to capture the signal
components to approximate the nonlinear function $f_i(x)$ which may
have a non-zero offset term $f_i(x_0)$ at $x_0$ (such a term has been
included in the bilinear model to be compensated, to inclrease the
model accuracy).

\bp
{\bf Advantage of online parameter estimation (adaptation) over
  offline parameter calculation}. The
adaptive algorithm (\ref{Thetak+1})--(\ref{Pk1}):
\beq
\hat{\Theta}(k+1) = \hat{\Theta}(k) - \frac{P(k-1) \zeta(k)
  \epsilon^T(k)}{m^2(k)},\;k = k_0, k_0+1, k_0+2,\ldots,
\eeq
with $\hat{\Theta}(k_0) = \Theta_0$ as a chosen initial estimate of
$\Theta$, where
\beq
P(k) = P(k-1) - \frac{P(k-1) \zeta(k) \zeta^{T}(k)
P(k-1)}{m^{2}(k)},\;k = k_0, k_{0}+1, k_0+2, \ldots,
\eeq
\beq
\epsilon(k) = \hat{\Theta}^T(k) \zeta(k) - y(k)
\eeq
\beq
m(k) = \sqrt{\rho + \zeta^{T}(k)P(k-1)\zeta(k)},\;\rho > 0
\eeq
generates the estimate $\hat{\Theta}(k)$ of the approximation model
parameter matrix $\Theta$ in (\ref{Theta}) for a linear model or in
(\ref{Thetab}) for a bilinear model, based on the minimization of
the cost function $J_3$ given in (\ref{Jrobot}) (which contains the
accumulative errors $(\hat{\Theta}^{T}(k) \zeta(\tau) -
y(\tau))^{T}(\hat{\Theta}^{T}(k) \zeta(\tau) - y(\tau))$):
\bea
J_3 = J_3(\hat{\Theta}) \ts = \ts \frac{1}{2} \sum_{\tau = k_0}^{k-1} 
\frac{1}{\rho} (\hat{\Theta}^{T}(k) \zeta(\tau) - y(\tau))^{T}
(\hat{\Theta}^{T}(k) \zeta(\tau) - y(\tau))\nn \\
\ts \ts +\frac{1}{2}\tr[(\hat{\Theta}(k) - \Theta_{0})^{T} P_{0}^{-1} (\hat{\Theta}(k) -
\Theta_{0})],\;P_{0} = P_{0}^{T} \left(\in R^{60 \times 60}\right) >
0,\;\rho > 0.
\eea
The calculation of $\Theta$ based on $\nabla f_i(x_0)$ from
(\ref{nablafi(x0)} and $H_i$ from (\ref{Hi0}) only captures those
local approximation terms, without a minimization mechanism.

\bp
{\bf Simulation study}. Our next research tasks will be focused on two areas of simulation:

\begin{quote}
$\bullet$ formulate the simulation systems, conduct the
related simulation studies, and present the simulation results, about
the parameter estimates $\hat{\Theta}(k)$ from (\ref{Thetak+1}),
estimation error $\epsilon(k)$ defined in (\ref{epsilon(k)}), model
validation error $\tilde{z}_{(2)}(k+1)$ defined in (\ref{tildez(2)}),
for both the linear model and bilinear model, to evaluate their
behaviors; and 

$\bullet$ evaluate the control system performance under a linear
programming control design using either a linear or bilinear adaptive
(online) Koopman system model.
\end{quote}

The outcomes of this simulation study will be reported in a companion
paper \cite{zth22}.

\section*{Acknowledgements}
The authors would like to thank the financial support from a Ford
University Research Program grant and the collaboration and help from
Dr. Suzhou Huang and Dr. Qi Dai of Ford Motor Company for this research.

\end{document}